\documentclass[unnumsec,webpdf,contemporary,large]{oup-authoring-template}%
\usepackage{bm}
\usepackage{graphicx}
\graphicspath{{Fig/}}
\onecolumn

\theoremstyle{thmstyleone}%
\theoremstyle{thmstyletwo}%
\theoremstyle{thmstylethree}%

\begin{document}

\journaltitle{---}
\DOI{DOI HERE}
\copyrightyear{---}
\pubyear{---}
\access{Advance Access Publication Date: ---}
\appnotes{Paper}

\firstpage{1}


\title[Mixed additive modelling of global alien species co-invasions of plants and insects]{Mixed additive modelling of global alien species co-invasions of plants and insects}

\author[1,$\ast$]{Martina Boschi}
\author[2]{Rūta Juozaitienė}
\author[1]{Ernst C. Wit}

\authormark{Boschi et al.}

\address[1]{\orgdiv{Faculty of Informatics}, \orgname{Univeristà della Svizzera italiana}, \orgaddress{\state{Lugano}, \country{Switzerland}}}
\address[2]{\orgname{Vytautas Magnus University}, \state{Kaunas}, \country{Lituania}}

\corresp[$\ast$]{\href{mailto:martina.boschi@usi.ch}{martina.boschi@usi.ch}}

\received{Date}{0}{Year}
\revised{Date}{0}{Year}
\accepted{Date}{0}{Year}


\abstract{Alien species refer to non-native species introduced by humans into an ecosystem, which can cause harm to the environment, economy, or human health. Although there is considerable literature on the subject, the presence of confounding factors has so far prevented a comprehensive picture of the relative importance of various drivers of such invasions. In this manuscript, we aim to develop and apply a general mixed additive relational event model to describe the pattern of global invasions of alien species. The diffusion of alien species can be regarded as a relational event, where the species -- the sender -- reaches a region -- the receiver -- at a specific time in history. We use the First Record Database, which contains all co-invasions by insects and plants between 1880 and 2005. A relational event model (REM) is employed to describe the underlying hazard of each species-region pair. Besides potentially time-varying, exogenous, and endogenous covariates, the mixed additive REM incorporates time-varying and random effects, allowing for taxa-specific baseline rates while accounting for the potential synergistic effect between plants and insects in the invasion process. Our efficient inference procedure relies on case-control sampling, yielding the same likelihood as that of a degenerate logistic regression. We propose fitting the mixed additive REM  via a generalised additive model with random effects as 0-dimensional splines. The resulting computational efficiency means that complex models for large dynamic networks can be estimated in seconds on a standard computer. Furthermore, we present a framework for testing the goodness-of-fit of our mixed additive REM for the invasions by vascular plants and insects by means of cumulative martingale-residuals. Implementation is performed  through the \texttt{R} package \texttt{mgcv}.}

\maketitle

\section{Introduction}\label{sec:introduction}

\textit{Alien species} refer to non-native species introduced by humans into a new ecosystem, successfully overcoming challenges, such as geographical barriers and sustaining reproduction in the new location. Although not all alien species become invasive, this phenomenon is now widely recognised as a significant and widespread threat due to the resulting environmental damage and costs involved. The detrimental consequences of invasive species include reducing biodiversity, damaging ecosystems, and impacting human health \citep{mcneely2001global,invasivespecies2010}. 
Understanding the mechanisms driving the dispersal of alien species and evaluating the impact of various factors on their rate of spread are crucial for effectively addressing this issue. While some invasions occur through diffusion processes, many introductions are deliberately facilitated by humans for commercial or aesthetic purposes \citep{pyvsek2020scientists}. Ecological, climatological, socioeconomic, historical, and geographical processes, along with species interactions, all contribute to this complex phenomenon.
Most of the aforementioned drivers are dynamic, changing over time and exhibiting significant fluctuations. For instance, international trade is widely acknowledged as one of the main factors driving biological invasions. Trade itself, and its influence on invasions, varies over time, as recent efforts such as international agreements and regular border surveillance aim to address the alien species invasions \citep{hulme2021unwelcome}.

So-called \emph{first records} (FRs)  are a way to investigate the spread of alien species. FRs are triplets, each of them consisting of the \emph{first year} in which a particular \emph{species} is detected in a particular \emph{region}. The Alien Species First Record Database is a global database that collects information on more than 47,000 FRs of established alien species \citep{seebens2018global}. Each record includes the year of the corresponding FR, as well as details about the species and the region involved. While information about the route, entry means, and occurrence frequency in a region would be undoubtedly valuable, this is typically unavailable. 

Several modelling approaches exist to describe the species diffusion processes. \textit{Species distribution models} (SDMs), a broad family of models widely used in ecology, seek to explain presence and abundance of species as a response to environmental changes, identifying regions at risk of invasion \citep{bellard2016major}. These models have also been expanded to the community level by considering multiple species simultaneously. However, the adequacy of SDMs must be evaluated carefully, particularly regarding the spatial and temporal congruence between the variables \citep{araujo2019standards, a2022species}.
Given the complexity and global scale of the invasion process, realistic modelling requires a framework that can transparently incorporate multiple, alternative scientific hypotheses. \citet{Juozaitiene2022} developed the first formulations of a \textit{relational event model} (REM) specific to the spread of alien species, considering the first record as a relational event whereby a species is ``connected'' to the non-native region at the moment of invasion. This explanatory model describes the possibly time-varying effect of time-varying covariates on the occurrence of the invasion events. However, the computational cost of their statistical inference procedure necessitated simplifications in the model formulation. Instead, we propose a general \textit{mixed-effect} \textit{additive} relational event model that incorporates non-linear, time-varying, and random effects, combined with an efficient estimation approach. This implies that we can consider larger relational event networks consisting of multiple interacting taxonomic groups of species. Specifically, we are interested in studying the \textit{joint invasion patterns} of \textit{vascular plants} and \textit{insects}. By modeling relational events involving species from different taxa jointly, we can estimate taxa-specific baseline rates while accounting for the influence of other taxa. Furthermore, this framework allows us to examine relationships among species across taxa, inferring how the species interactions influence the invasion process.

In section \textit{Data Sources} we describe the characteristics of the Alien Species First Record database and the additional sources of data that are employed in this study. The section entitled \textit{Generative Species Invasion Model} aims at defining the mixed additive relational event model involving time-varying covariates, time-varying effects and random effects. The following section  proposes an efficient inference method for the model, as well as an associated goodness-of-fit technique. After showing its accuracy in a target simulation study, we analyse the global joint invasion patterns of insects and vascular plants. The code used in this study is publicly available on GitHub at \url{https://github.com/martinaboschi/alienspecies.git}. We conclude by discussing how our work relates to recent literature on invasion dynamics and species co-invasion structures, followed by a consideration of methodological limitations and potential future research paths.

\section{Data Sources} \label{sec:materials}

In this section we describe the various data sources and records we consulted for our study. Our aim is to infer a parsimonious quasi-generative model for the spread of insects and vascular plants in recent history. Specifically, this model aims to describe the mechanisms of invasion using the available information, which is presented in this section. First, we describe the information we have on the temporal establishment of certain species in specific regions. Then we focus on information relating to potential drivers of the invasion process. Additionally, as a starting point of our analysis, we consider the native ranges at 1880. These native ranges are extremely important as all future invasions start from the current native ranges.

\subsection{Alien Species First Record Database}\label{subsec:ASFR}

We use version 2.0 of the \textit{Alien Species First Record} (ASFR) Database \citep{seebens2017no, seebens2018global} consisting of  47,542 invasions involving 16,922 species across 275 regions. Data is available at \url{https://zenodo.org/record/4632335}. Each first record includes the year a species was first documented in a specific region before becoming established there. Regions refer to either countries or islands that are geographically distant from the nation to which they politically belong. A first record also contains information on the source that reported the invasion. 

Although the ASFR database contains recorded invasions from 7,000BC to the present day, our analysis focuses on the period between 1880 and 2005. The lower limit of 1880 was chosen because it was only in the 19th century that the reporting of invasions became more systematic. Furthermore, invasion curves were nearly flat before 1800, and the relevance of alien species invasions has almost certainly risen with the onset of \textit{globalisation} \citep{seebens2021around}. Despite having data up to the present, an upper limit of 2005 was chosen to minimise potential delays in recording invasions.

Our analysis focuses on two of the most frequently cited land-based taxa, namely insects (19\%) and vascular plants (52\%). 
It is well known that certain insects, such as various pollinators, directly influence the spread of plants \citep{russo2016positive}, a relationship sometimes referred to as mutualism. \textit{Mutualism} can be considered as a form of cooperation between species and is recognised as playing a role in facilitating plant introductions \citep{richardson2000plant}. Other potential relationships, such as commensalism, parasitism, or predation, may also exist. Our aim is to study the co-invasion of plants and insects, with a focus on identifying the nature of their relationships.

\subsection{Data on Potential Drivers of Alien Species Invasions}\label{subsec:covariate}
Global alien species invasion is a complex process that involves a myriad of factors, ranging from geo-economical and geo-political drivers to climatic and geographical elements. We restricted our attention to those factors that have been hypothesised as important and for which relatively reliable information was available. Table~\ref{tab:covariates} describes  the set of drivers that are considered in our work, their possible temporal variability, and sources.

\begin{sidewaystable}
	\centering
		\begin{tabular*}{\textwidth}{@{\extracolsep{\fill}}p{0.15\textwidth}|p{0.05\textwidth}|p{0.3\textwidth}|p{0.17\textwidth}|p{0.15\textwidth}|p{0.10\textwidth}@{\extracolsep{\fill}}}
		\toprule
		\textbf{Covariate Name} & \textbf{Symbol} & \textbf{Definition} & \textbf{Information} & \textbf{Type} & \textbf{Source} \\
		\hline
		\textbf{\textit{distance}} & $ d_{sr}(t) $ & Logarithm of the distance from the region $ r $ and the nearest region invaded by species $ s $ by time $ t $. & Distance between two countries is defined as the distance between their closest borders. & Time-varying covariate with time-varying effect & \citep{distancesource} \\
		\hline
		\textbf{\textit{trade}} & $ tr_{sr}(t) $ & Logarithm of the sum of annual trade flows (in current United States dollars) between region $ r $ and other countries that have been invaded by species $ s $ before time $ t $. & This variable shows several missings: our imputation method is described in Section~\ref{subsec:covariate}. & Time-varying covariate with time-varying effect & \citep{barbieri2009trading} \\
		\hline
		\textbf{\textit{climatic dissimilarity}} & $ dt_{sr}(t) $ & Minimum difference in near-surface air temperature (in absolute value) between region $ r $ and other countries that have been invaded by species $ s $ before time $ t $. & --- & Time-varying covariate with fixed effect & \citep{watanabe2011miroc} \\
		\hline
		\textbf{\textit{agricultural land-coverage}}& $ l_r(t) $ & Sum of cropland and pasture proportions in the country $ r $ at time $ t $. & --- & Time-varying covariate with time-varying effect & \citep{landsource} \\ 
		\hline
		\textbf{\textit{urban land-coverage}} & $ u_{r}(t) $ & Urban area proportion in the country $ r $ at time $ t $. & --- & Time-varying covariate with fixed effect & \citep{landsource} \\ 
		\hline
		\textbf{\textit{colonial ties}} & $ k_{sr}(t) $ & Presence of the species $ s $ at time $ t $ within the colonial power to which region $ r $ has belonged. & Each country is characterised either as independent or by a colonial empire it belonged to. & Time-varying covariate with fixed effect & \citep{colonialsource} \\ 
		\botrule
	\end{tabular*}
	\caption[Potential drivers of alien species invasions]{\label{tab:covariates} Potential drivers of alien species invasions. This table is an adaptation of Table 1 and 2 in \citet{Juozaitiene2022}. It reports the notation, definition, effect type, and data source for each driver. Additional information may be included.}
\end{sidewaystable}

Invasion events over long geographical distances are relatively uncommon \citep{Trakhtenbrot2005}. In order to evaluate the precise role of distance, we consider the shortest distance among the regions in which the species is already present at that time. \textit{distance} is computed referring to the closest borders, resulting in a value of zero for neighbouring regions. Source data for this driver consists of the \textsf{R} package \textsf{geosphere} \citep{distancesource}. In order to avoid issues with outliers, distances are log-transformed. 

In the existing literature, international trade has been acknowledged as a key factor for explaining the spread of alien species; the value of import commodities is a common proxy for the rate of alien species introductions \citep{seebens2018global}. Sometimes, the two terms are even used interchangeably \citep{hulme2021unwelcome}. 
Source information comes from \citet{barbieri2009trading}, and reports trade flows among countries. Trade values are not available for every pair of countries in each year considered. \citet{Juozaitiene2022} proposed an imputation method for the source data: when gaps emerge at the beginning of the observation period, they are replaced with zero; instead, gaps in intervening years of already recorded trade are imputed according to a log-linear model in case of growing trade. If a gap occurs at the end of the observation period, then the last available trade is carried forward. We define the quantity of interest, \textit{trade}$(s,r)$, as the yearly commerce between already-invaded territories by $s$ and the region $r$ (recorded as bilateral trade flow in current United States dollars). 
As trade varies by orders of magnitude, the values are log-transformed in the analysis. 

\citet{bellard2016major} report the climate as a relevant driver for invasions. Our analysis relies on the inferred yearly values of near-surface air temperature from \citet{watanabe2011miroc}. Various studies have highlighted the significance of average annual temperature as a key component of climatic conditions  \citep{seebens2018global,colling2025ninety,barni2012establishing}. While precipitation patterns exhibit greater variability than temperature under climate change \citep{Finch2021}, temperature remains strongly correlated with precipitation ($r=0.8$) \citep{colling2025ninety}. Furthermore, \citet{barni2012establishing} reported significant multicollinearity among climate variables and identified minimum temperature, mean temperature, and the heat sum for the spring season as the most important climatic factors. We introduce \textit{climatic dissimilarity} for each country and species as the minimal temperature difference relative to all other countries where that species was already present before the year of interest. 

The proportion of cropland, pasture and urban areas in land coverage \citep{landsource} are employed to assess which land-use features encourage introductions and establishment of alien species. \citet{seebens2018global} suggested their role in the variation in first records for several taxonomic groups. In our analyses we employ the variables \textit{agricultural land-coverage} and \textit{urban land-coverage}. 

Colonial expansion, particularly European colonialism by the British, Spanish, Portuguese and Dutch, has been recognized as a factor in the dispersion of alien species \citep{dyer2017global, lenzner2022naturalized}. Data on colonial ties are available from the COLDAT infrastructure \citep{colonialsource}, which reports the presence and the starting and ending date of colonial powers\footnote{Belgium, Britain, France, Italy, Germany, Netherlands, Portugal, and Spain are the reported colonial powers.} in the regions mentioned in \citet{barbieri2009trading}. In our analysis, each region either refers to the colonial power it belonged to or is classified as independent. 
For each species-region-year triplet, the covariate \textit{colonial ties} is an indicator function of whether the species is already present in the area associated with the colonial power to which the region belongs.

\subsection{Native Range}
A species' \textit{native range} (NR) is the collection of areas where it is indigenous. Slightly more liberally, we refer to NR as the set of sites where a species was already present before start of the analysis period, which in this context is 1880. This notion is relevant both in ecological terms and statistical terms. First, the ASFR database does not provide information on the origin of the species, but only the region it is invading. Knowledge of its NR allows us to identify which possible parts of the world this species hails from. Additionally, fascinating relationships between native and invaded environments may be examined \citep{hejda2015native}. Secondly,  several of the factors mentioned in the previous section are \emph{endogenous} --- this means that these factors arise from the sequence of previously occurred invasions. For each species-region-year triplet, we need to know where the species was already present prior to that specific year. In this sense, the notion of NR is necessary statistically to compute the covariate values. \citet{van2019global} and CABI Invasive Species Compendium ({\footnotesize \url{https://www.cabi.org}}, accessed 15.07.2016) are used as sources that describe the NRs of vascular plants and insects. Additionally, the ASFR invasions before 1880 were used to supplement the NRs.

Table~\ref{tab:datastructure} reports the structure of the analyzed data, including the 13,094 invasion events between 1880 and 2005 reported in the ASFR database, together with the number of species and regions involved. The number of already occurred species-regions dyads, recorded in NR, is also reported.

\begin{table*}[t]
	\centering
	\begin{tabular*}{\textwidth}{@{\extracolsep{\fill}}ccccc@{\extracolsep{\fill}}}
		\toprule
		Taxonomic Group & No. of FR before 1880 & No. of FR 1880-2005 & No. of species & No. of regions \\ 
		\midrule
		Insects & 1098 & 586 & 114 & 159 \\ 
		Vascular plants & 60448 & 12508 & 3921 & 120 \\ 
		Insects and Plants & 61546 & 13094 & 4035 & 188 \\ 
		\botrule
	\end{tabular*}
	\caption[Main features of the final data]{\label{tab:datastructure} Main features of the final data structure, including overall cardinality of native range sets, number of invasion events from the ASFR database between 1880 and 2005 involving plants and insects, and number of species and regions involved.}
\end{table*}

\section{Generative Species Invasion Model}\label{sec:model}

An alien species invasion by species $s$ of a region $r$ in year $t$ can be considered a \textit{relational event}. A relational event, involving a sender $s$ interacting with a receiver $r$ at time $t$, can be expressed as a triplet $e=(s,r,t)$. 
The FR \textit{invasion process} (IP) can be modelled as a \textit{marked point process} (MPP) $ \{[t_k, (s_k,r_k)]; k \ge 1 \}$, where species-region dyad can be considered a mark on the event process. FR sequences can thus be expressed as,
\begin{displaymath}
	\begin{aligned}
		\mathcal{E} = \{ e_{k}|e_{k} = (s_{k}, r_{k}, t_{k}) \subseteq \mathcal{S} \times \mathcal{C} \times T, \quad k=1,...,n \} 
	\end{aligned}
\end{displaymath}
where $\mathcal{C}$ is the set of global regions, $\mathcal{S}$ is the collection of plants and insects, and $T$ is the period of interest. We focus on $T=[1880,2005]$, the years between $ 1880 $ and $ 2005$. We associate with the IP a \textit{counting process} (CP) $ \{N_{sr}~\mid~s\in \mathcal{S}, r \in \mathcal{C}\}$, counting the number of marks $(s,r)$ in $[1880,t]$,
\begin{equation}\label{1}
	N_{sr}(t) = |\{\text{invasions of species $s$ in region $r$ by time $t$} \}|
\end{equation}
FR are \textit{non-recurrent events}: if an introduction $(s,r)$ is observed at time $t$, the dyad is not at risk to occur anymore. Therefore, the counting process associated to the IP takes values $0$ or $1$, and is thus almost surely finite. CP $N_{sr}$ is adapted with respect to the increasing \textit{filtration} $\mathbb{F} = \{\mathcal{F}_t\}_{t \ge 1880}$. At time $t$, we incorporate into $\mathcal{F}_{t^-}$ the history of the process prior to $t$. Conditionally on their history and associated covariate process, alien species invasions are assumed to occur \textit{independently} \citep{butts_4_2008, perry_point_2013, vu2017relational}. Furthermore, we assume no simultaneous events can occur, and that the CP process starts at 1880, i.e., $N_{sr}(1880)=0 \quad \forall s \in \mathcal{S}, r \in \mathcal{C}$. With these properties and its non-decreasing nature, CP is a continuous-time submartingale and, as such, it can be decomposed in accordance to the \textit{Doob-Meyer theorem}:
\begin{displaymath}
	N_{sr}(t)= \Lambda_{sr}(t) + M_{sr}(t)
\end{displaymath}where $ M_{sr}(t) $ is a continuous-time \textit{martingale}, and \textit{cumulative hazard} $\Lambda_{sr}(t) = \int_{1880}^t \lambda_{sr}(u)  \mathrm{d} u$ is a \textit{predictable} increasing process. The hazard $\lambda_{sr}(t)$ is measurable with respect to ${\mathcal{F}}_{t^-} \forall t \ge 0$. 

The aim of this work is to model the \textit{intensity function} $ \lambda_{sr}(t) $ as a function of possibly time-varying socio-economical, ecological and geographical drivers, including the statistical significance, direction, and size of their effects. 
We aim to model the intensity function of the CP $ \{N_{sr}\}$ as mixed-effect additive relational event model, including time-varying covariates with potential time-varying effects and random effects: 
\begin{equation}\label{3}
	\begin{aligned}
		\begin{aligned}
			\lambda_{sr}(t|{\mathcal{F}}_{t^-};\bm{\beta}, \bm{\theta})  = \lambda_{0a}(t)  \exp{\left[ \bm{\beta}(t)' \bm{x}_{sr}(t) + \bm{b}' \bm{z}_{sr}(t) \right]} \\ 
			\bm{b} \sim \mathcal{N}(\bm{0}, \Sigma(\bm{\theta}))
		\end{aligned}
	\end{aligned}
\end{equation}
where: 
\begin{itemize}
	\item $ \lambda_{0a}(t)$ is a non-negative \textit{stratified baseline} intensity function; it captures the residual hazard that is not explained by the drivers that are included in the model formulation. It is permitted to vary in the different strata $a$. In our analysis, the baseline varies between two taxonomic groups: vascular plants and insects.
	\item $ \bm{x}_{sr} $ and  $ \bm{z}_{sr} $ are left-continuous, adapted, and thus predictable and locally bounded \textit{covariate processes}. 
	\item $ \bm{\beta}$ are fixed, potentially time-varying, effects.
	\item $ \bm{b} $ are \textit{random frailties}, capturing additional heterogeneity.
\end{itemize} 
The choice of explanatory variables to be included in the model specification is non-trivial. We discuss our approach in section \textit{Analysis of Plant and Insect Co-invasions}.

\subsection{Fixed, possibly Time-Varying, Effects}
The covariate process $\bm{x}$ can consist of exogenous and endogenous variables. \textit{Exogenous} covariates are external to the process, whereas endogenous covariates are functions of the IP itself. Due to the intricate nature of the IP process, all variables described in Table~\ref{tab:covariates} are endogenous. The impact of these time-varying drivers on the rate of occurrence may either be assumed to be fixed or allowed to vary over time. The nature of the effect for each variable is reported in Table~\ref{tab:covariates}.
Time-varying effects for covariate $j$ are defined as $\bm{\beta}^j(t)=\sum_i \bm{\beta}^{j}_i \bm{g}^j_i(t)$. A possible option for $\bm{g}$ is a set of \textit{radial basis functions}. The resulting \textit{thin plate splines} are extremely flexible but involve a sizeable number of parameters. A \emph{thin plate regression spline} is a low-rank approximation of thin plate splines that can be incorporated into a wide range of models \citep{tprs}. We consider the latter.

\subsection{Random Effects}
One of the goals of the study is to understand if heterogeneity of species invasiveness and region invasibility plays a significant role in the invasion process. Additionally, we want to explore the overall co-invasion patterns of species by examining if the presence of a species in a particular region affects the rate of invasion by other organisms in that region. We thus include two kinds of random effects: \textit{monadic random intercepts} capture heterogeneity of species and regions, whereas \textit{dyadic random intercepts} model heterogeneity in the co-invasion of species pairs. It considers pairs consisting of the currently invading species and the most recent species to enter the area. Due to the difference in the order of magnitude of the number of insects and plants, we decided to model the species invasiveness random effect with between-strata heteroscedasticity.

\section{Efficient Inference Method}\label{sec:inference}

We consider the relational event sequence $\mathcal{E}$, consisting of $n$ relational events. The estimation procedure for the fixed and random effects in event history models typically relies on the \textit{partial likelihood} (PL), which treats the baseline hazard function $\lambda_{0a}(t)$ as a nuisance parameter. Although computationally more efficient than full maximum likelihood estimation (MLE) for large event sequences with time-varying covariates, the method becomes computationally prohibitive as the risk set scales as $O(|\mathcal{S}|\times |\mathcal{C}|)$. For this reason, we focus on a sampled version of the partial likelihood, whose runtime complexity does not change with the size of the sender or receiver sets.

\subsection{Case-Control Partial Likelihood Inference via GAMs}\label{subsec:inference}

\textit{Nested case-control} (NCC) sampling \citep{borgan1995methods, lerner2020reliability} considers, for each event $(s, r)$ in $\mathcal{E}$, a reduced risk set composed of the event and $m-1$ non-events, sampled according to a given probability distribution $ \pi_{t}(\cdot|sr)$. We consider the case in which $m=2$, i.e., at each time $t$ the \textit{sampled risk set} $\bm{sr}$ consists of the \textit{event dyad} $(s,r)$ and one \textit{non-event dyad} $(s^\ast, r^\ast)$ randomly sampled from the complete \textit{risk set} $\mathcal{R}(t)$, which consists of the pairs of species and regions that could be observed at time $t$.

We define a new MPP $ \{\left[t_k, (s_k, r_k, SR_{t_k}) \right]; k = 1,\ldots,n \} $ where $SR_{t_k}$ is the sampled risk set at time $t_k$. The \textit{marked space} of the MPP is given by $E = \{ (s,r, \bm{sr})~\mid~s \in \mathcal{S}, r \in \mathcal{C}, \bm{sr} \in \mathcal{P}_{sr} \}$, where $\mathcal{P}_{sr}$ is the subset of the \textit{power set} $\mathcal{P}$ of all dyads that contains the event $(s,r)$; in particular, when $m=2$,  it consists of $|\mathcal{R}|-1$ sets consisting of the event $(s,r)$ and one other dyad at risk, i.e., $\mathcal{P}_{sr} = \{[(s,r),(s^\ast,r^\ast)]\mid(s^\ast,r^\ast)\in\mathcal{R}\}$. Also to this MPP, we can associate a CP,
\begin{equation}\label{6}
	N_{(s,r,\bm{sr})}(t) = \sum_{t_k \le t} 1_{\{ (s_k,r_k,SR_{t_k})=(s,r,\bm{sr})\}}
\end{equation}
where the original CP in \eqref{1} can be retrieved as $N_{sr} = \sum_{\bm{sr} \in \mathcal{P}_{sr}} N_{(s,r,\bm{sr})}$. 
With the new CP we associate a new filtration $\mathcal{H}_t = {\mathcal{F}}_t \cup \sigma\{SR_{t_k}; t_k \le t\}$, that consists of the cohort history augmented with the risk set sampling information. We assume \textit{independent sampling}, meaning that sampling probabilities do not depend on the event risk. Under this assumption, the intensity process of the CP $N_{sr}$ is adapted not only to $\mathbb{F}$ but also to $\mathbb{H} = \{ \mathcal{H}_t \}_{t \ge 0}$ \citep{borgan2015using}.

We can decompose the intensity process of the CP $N_{(s,r,\bm{sr})}(t)$ in two different ways: 
\begin{displaymath}
	\begin{aligned}
		\lambda_{(s,r,\bm{sr})}(t) &= \left\{ \begin{array}{l}
			\lambda_{sr}(t) \pi_t(\bm{sr}|(s,r)) \\
			\lambda_{\bm{sr}}(t) \pi_t((s,r)|\bm{sr})
		\end{array} \right.
	\end{aligned}
\end{displaymath}
On the one hand, in the case of NCC sampling with $ m=2 $, $ \pi_t(\bm{sr}|s,r) $ is taken as equal for all sets $ \bm{sr} \in \mathcal{P}_{sr}(t)$,
\begin{equation}\label{7}
	\pi_t(\bm{sr}|(s,r)) = \frac{1}{|\mathcal{R}(t)|-1} \cdot I_{\{(s,r) \in \bm{sr}\}}. 
\end{equation}
In the case of a stratified REM, such as we consider in \eqref{3}, the sampling of the non-events is constrained to the stratum of the event \citep{borgan1997risk}. 

On the other hand, the probability of the dyad $(s,r)$ occurring at $t$ given that some $\bm{sr}$ is sampled is given as:
\begin{equation}\label{8}
	\pi_t((s,r)|\bm{sr}, \mathcal{H}_{t^-}) = \dfrac{\lambda_{(s,r,\bm{sr})}(t|\mathcal{H}_{t^-})}{\lambda_{\bm{sr}}(t|\mathcal{H}_{t^-})}
\end{equation}

Due the conditional independence of events given their previous history, the joint product of probabilities in \eqref{8}, when $m=2$, yields a reduced partial likelihood, $\bm{\mathcal{L}}_S$, the \emph{sampled PL},
\begin{equation}\label{9}
		\bm{\mathcal{L}}_S(\bm{\beta}, \bm{\theta}) =  
		\prod_{k=1}^{n}  \Big{\{}1+\exp{\left[- \left(\bm{\beta}(t_k)' \cdot \Delta\bm{x}_k + \bm{b}' \cdot \Delta\bm{z}_k \right) \right]}\Big{\}}^{-1}.
\end{equation}
Given the sampled non-event $(s_k^\ast, r_k^\ast)$, \textit{covariate differences} are defined as follows,
\begin{equation*}
	\begin{aligned}
		& \Delta\bm{x}_k= \bm{x}_{s_k r_k}(t_k) - \bm{x}_{s^\ast_k r^\ast_k}(t_k)\\ 
		& \Delta\bm{z}_k= \bm{z}_{s_k r_k}(t_k) - \bm{z}_{s^\ast_k r^\ast_k}(t_k)\\ 
	\end{aligned}
\end{equation*}
Expression \eqref{9} is not only computationally less expensive but also corresponds to the likelihood of a mixed additive \textit{logistic regression} model without intercept, where the observed responses $y_1,\ldots,y_n$ are all successes, and the covariates are defined as the covariance difference for the pairs of events and sampled non-events,
\begin{equation}\label{10}
	\begin{aligned}
		&Y_k|\Delta\bm{x}_k, \Delta\bm{z}_k, \bm{b}\stackrel{\mbox{iid}}{\sim} \text{Bernoulli}\left( \pi_k \right),~~ k=1,\ldots,n  \\
		& \mbox{logit}(\pi_k)= \bm{\beta}(t_k)' \cdot \Delta\bm{x}_k + \bm{b}' \cdot \Delta\bm{z}_k   \\  
	\end{aligned}
\end{equation}
Since  \eqref{9} may be expressed as the likelihood of \eqref{10}, we estimate the mixed-effect additive relational event model by fitting a \textit{generalised additive mixed model} (GAMM). We include the time-varying effects as thin plate regression splines. Random effects can be efficiently estimated as smooth terms of dimension $0$, with basis functions taking the value $1$ when the level of the random factor is present and $0$ otherwise. Each of the smooth terms involves a penalisation term to the sampled PL in \eqref{9}. Specifically, normally distributed random effects arise from a penalty that consists of an identity matrix of dimension equal to the number of levels of the corresponding random factor \citep{pedersen2019hierarchical}. The inference technique is implemented by using the \textsf{R} package \textsf{mgcv} \citep{tprs, woodAIC, wood_generalized_nodate}.

\subsection{Non-Parametric Estimation of the Baseline Hazard}
The \textit{cumulative baseline function} $\Lambda_{0}(t) = \int_{t_0}^t \lambda_{0}(u) \mathrm{d}u$ may be estimated non-parametrically.  \citet{borgan1995methods} and \citet{borgan1997risk} propose an adaptation of the \textit{Breslow estimator} for sampled cohort data. Consider the expected fitted linear predictors $ \hat\eta_{sr}(t) = \hat{\bm{\beta}}(t)' \cdot \bm{x}_{sr}(t) + \mathbb{E}[\bm{b} \mid \mathcal{E}]' \cdot \bm{z}_{sr}(t) = \hat{\bm{\gamma}}' \bm{h}_{sr}(t)$, where $\hat{\bm{\gamma}}$ includes the GAMM fitted parameter and $\bm{h}_{sr}(t)$ the corresponding model matrix. Then, the Breslow estimator can be written as  
\begin{equation}\label{11}
	\widehat{\Lambda}_0(t|{\mathcal{H}}_{t^-}) = 
	\sum_{t_k \le t} \dfrac{1}{\sum_{(s,r) \in SR_{t_k}} \exp{\left[\hat\eta_{sr}(t_k)\right]} \cdot \omega_{sr}(t_k,SR_{t_k})}
\end{equation}
where,
\begin{equation*}
	\omega_{sr}(t,\bm{sr}) = \dfrac{\pi_t(\bm{sr}|(s,r))}{|\mathcal{R}(t)|^{-1} \sum_{(s^\ast, r^\ast) \in \bm{sr}} \pi_t(\bm{sr}|(s^\ast,r^\ast))}
\end{equation*}
Since we consider matched NCC with $m=2$ with a stratification $a$ of the species population into $a_1 =$ plants and $a_2 =$ insects, the weights in the baseline hazard estimate $\widehat{\Lambda}_{0a}$ simplify as $\omega_{sr}(t,\bm{sr}) = |\mathcal{R}_a(t)|/2$. Alternatively, the baseline hazard in this setting may be estimated using shifted nested-case control sampling, as described in \cite{lembo2024relational}.

\subsection{Goodness of Fit Evaluation}\label{subsec:goodnessfit}
\textit{Goodness of fit} (GOF) evaluation for relational event models is still an underexplored subject \citep{brandenberger_predicting_2019}. Informal approaches for the Cox proportional hazards model, including the evaluation of Schoenfield, deviance, and martingale residuals have been extended to relational event models \citep{Juozaitiene2022}. Instead, \citet{brandenberger_predicting_2019} has proposed an approach comparing the data to simulated draws from the fitted model. Another simulation-based method for assessing the modelling of auxiliary statistics under the fitted model has been recently proposed by \cite{amati2024goodness}. The guiding principle behind these approaches is that the former should resemble the latter, if the fit of the model is good. Both methods are computationally expensive, and rely on several assumptions that are difficult to check. 

In this manuscript, we adapt a third approach, originally proposed in the survival literature.  \citet{lin1993checking, borgan2015using} introduce a temporal process consisting of cumulative sums of martingale residuals for the Cox proportional hazards model. Under the null hypothesis, this sum has a known asymptotic distributional behaviour. However, given that the presence of time-varying effects in our model formulation violates the proportional hazard assumption, our proposal further relies on the goodness-of-fit tools presented in \citet{marzec1997}. Specifically, we rely on a zero-mean Martingale-residual type process $G[\hat{\bm{\gamma}}, \cdot|\mathcal{E}]$, defined as a weighted cumulative sum of \textit{Martingale residuals} $\widehat{M}_{sr}=N_{sr}(t) - \widehat{\Lambda}_{sr}(t)$, for any statistic of interest $\phi$, and evaluated at $n$ equally spaced points $u \in [0,1]$:
\begin{equation}\label{13}
	\begin{aligned}
		G[\hat{\bm{\gamma}}, u|\mathcal{E}] &= \sum_{k \le \lfloor nu \rfloor}
		w_{s_kr_k}(t_k) \cdot \nabla \widehat{M}_{s_kr_k}(t_k) \cdot
		\phi_{s_kr_k}(t_k)    \\ &=  \sum_{k \le \lfloor nu \rfloor} \left[
		w_{s_kr_k}(t_k) \cdot \phi_{s_kr_k}(t_k)- \dfrac{\Phi^{(0)}_{\bm{sr}}[\hat{\bm{\gamma}},t_k]}{S^{(0)}_{\bm{sr}}[\hat{\bm{\gamma}},t_k]}\right]
	\end{aligned}
\end{equation}
where $w_{sr}(\cdot) $ is any weight function that assumes values in $(0,1]$,  $\nabla \widehat{M}_{s_kr_k}(t_k) = \left[1 - \nabla \widehat{\Lambda}_{s_kr_k}(t_k) \right]$ is the $k$th increment in the Martingale residuals process, and $\nabla \widehat{\Lambda}_{s_kr_k}(t_k)$ is the $k$th increment in the cumulative intensity process due to observation $k$. We also define the following quantities:
\begin{equation*}
	\begin{aligned}
		\Phi^{(0)}_{\bm{sr}}[\bm{\gamma},t] &= \sum_{sr \in \bm{sr}} \phi_{sr}(t) \cdot \exp{\left[ {\bm{\gamma}}' \bm{h}_{sr}(t) \right]} \cdot \pi_{t}(\bm{sr}|sr) \\
		S^{(0)}_{\bm{sr}}[\bm{\gamma},t] &= \sum_{sr \in \bm{sr}} \exp{\left[ {\bm{\gamma}}' \bm{h}_{sr}(t) \right]} \cdot \pi_{t}(\bm{sr}|sr) 
	\end{aligned}
\end{equation*}

The curve $G$ measures the difference between the observed statistic $\phi$ and its expected value across time. In particular, we are interested in testing whether the covariates are included appropriately in the model, i.e., $\phi=x^j$. In this case, $G[\hat{\bm{\gamma}}, \cdot|\mathcal{E}]$ can be shown to be a bridge process with $G[\hat{\bm{\gamma}}, 0|\mathcal{E}]=G[\hat{\bm{\gamma}}, 1|\mathcal{E}]=0$.  Given an estimate for the variance of an individual contribution to process $G[\hat{\bm{\gamma}}, \cdot|\mathcal{E}] $, named $\hat{J}_{G[\hat{\bm{\gamma}}]}$, it is possible to implement a formal, \textit{Kolmogorov-Smirnov} (KS) type statistical test:
\begin{equation}\label{15}
	KS = \sup{\{|\hat{J}_{G[\hat{\bm{\gamma}}]}^{-\frac{1}{2}} \times n^{-\frac{1}{2}} \times G[\hat{\bm{\gamma}}, u]|: u \in [0,1]\}} 
\end{equation}
When $\phi$ is univariate $x^j$, the p-value of the statistical test can be directly found by evaluating the Kolmogorov cumulative probability distribution at the observed value of the statistic $KS$. 

In more complex scenarios, such as the one involving time-varying or random effects, we need to consider a multivariate process. Particularly, when the covariate has a time-varying effect, for each time point, we have $q$ elements in the model matrix that refer to it, consisting of the evaluation of $q$ basis functions of time multiplied by the value of the $j$th covariate, both evaluated at the time of interest. Instead, to inspect each random factor, we need to simultaneously consider all the elements of the model matrix $z_j$s referring to the presence or absence of the related level. An additional element of complexity in these scenarios is the penalization term that is involved in the mixed additive likelihood. 

Let $ \bm{\phi}_{sr}(t) = \bm{h}_{\bm{i}, sr}(t) $ be a generic subset of $q$ elements in the model matrix, indexed by $\bm{i}$ and evaluated at time $t$. When a penalization term $P^{\lambda}(\bm{\hat\gamma})$ is included in the log-likelihood, the score vector evaluated at the penalised MLE equals the derivative of the penalty, i.e., $\nabla \bm{\ell}_S(\bm{\hat\gamma}) = \nabla P^{\lambda}(\bm{\hat\gamma})$, where $\bm{\ell}_S = \log{\bm{\mathcal{L}}_S}$. In order to define a bridge process that returns to zero, we recenter each $q$-dimensional individual term $\bm{G}^{\bm{i}}_{s_kr_k}(u)$ to have zero mean, 
\begin{equation*}
	\bm{G}^{\bm{i}}[\hat{\bm{\gamma}}, u|\mathcal{E}] = \sum_{k \le \lfloor nu \rfloor} \left[ \bm{G}^{\bm{i}}_{s_kr_k}[\hat{\bm{\gamma}}, t_k] -\frac{\nabla_{\bm{i}} P^{\lambda}(\bm{\hat\gamma})}{n} \right], \qquad \nabla_{\bm{i}} P^{\lambda}(\bm{\hat\gamma}) = \left. \frac{\partial P^{\lambda}(\bm{\gamma})}{\partial\bm{\gamma_{\bm{i}}}} \right|_{\bm{\gamma}_{\bm{i}} = \bm{\hat\gamma}_{\bm{i}}}
\end{equation*}
As an estimate for the covariance function of the individual contribution to the multivariate process we rely on the empirical variance covariance function,
\begin{equation*}
	\hat{\bm{J}}_G = n ^{-1} \times \sum_{k=1}^{n} \bm{G}^{\bm{i}}_{s_kr_k}[\hat{\bm{\gamma}}, t_k] \bm{G}^{\bm{i}}_{s_kr_k}[\hat{\bm{\gamma}}, t_k]'
\end{equation*}
If $\bm{G}^{\bm{i}}[\hat{\bm{\gamma}}, \cdot|\mathcal{E}]$ is the unpenalised score vector, one can use instead the \textit{observed Fisher information matrix}. It follows that the scaled score vector,
\begin{equation*}
	\hat{\bm{M}}^{\bm{i}}[\hat{\bm{\gamma}}, \cdot|\mathcal{E}] = 
	\hat{\bm{J}}_G^{-\frac{1}{2}} \times n^{-\frac{1}{2}} \times \bm{G}^{\bm{i}}[\hat{\bm{\gamma}}, \cdot|\mathcal{E}]
\end{equation*}
converges to a a multivariate Brownian bridge $\bm{Z}^0$ under the assumption that the model is correctly incorporating the covariate \citep{hjort2002}. To test the fit of a covariate whose elements in the model matrix correspond to $\bm{h}_{\bm{i}, sr}$, we propose the following test-statistic,
\begin{equation}\label{eq-KS}
	\textmd{KS}^{\bm{i}} = \sup_{u\in[0,1]}{\lVert \hat{\bm{M}}^{\bm{i}}[\hat{\bm{\gamma}}, u|\mathcal{E}] \rVert^2}
\end{equation}
Under the assumption of adequacy of the model formulation, this statistical test converges to the supremum of a $q$-dimensional \textit{Brownian bridge} $\sup_{u\in[0,1]}{\lVert  \bm{Z}^0(u) \rVert^2}$, as $n \rightarrow \infty$. The quantity $\sup_{u\in[0,1]}{\lVert  \bm{Z}^0(u) \rVert^2}$ can be empirically simulated and the p-value of the test is estimated as the proportion of simulated statistics larger or equal to the observed one. 

Finally, an \textit{omnibus test} allows for testing the overall fit of the model formulation, 
\begin{equation}\label{eq:omnibus_test}
	T_g = \max_{l=1, \ldots, L} \left\{ \frac{\max_{u\in [0,1]} \lVert \hat{\bm{M}}^{\bm{i}_l}[\hat{\bm{\gamma}}, u] \rVert^2}{|\bm{i}_l|} \right\} 
\end{equation}
where \( L \) represents the total number of covariates in the model, and \( |\bm{i}_l| \) represents the number of elements in the score vector that refer to the \( l \)-th group. The empirical p-value related to this omnibus test can be obtained by simulation, relying on the asymptotic convergence of each \( \hat{\bm{M}}^{\bm{i}_l}[\hat{\bm{\gamma}}, \cdot] \) to a \( |\bm{i}_l| \)-dimensional Brownian bridge.


\begin{figure}[!t] 
	\centering
	\begin{tabular}{cc}
		\includegraphics[height=6.7cm]{ 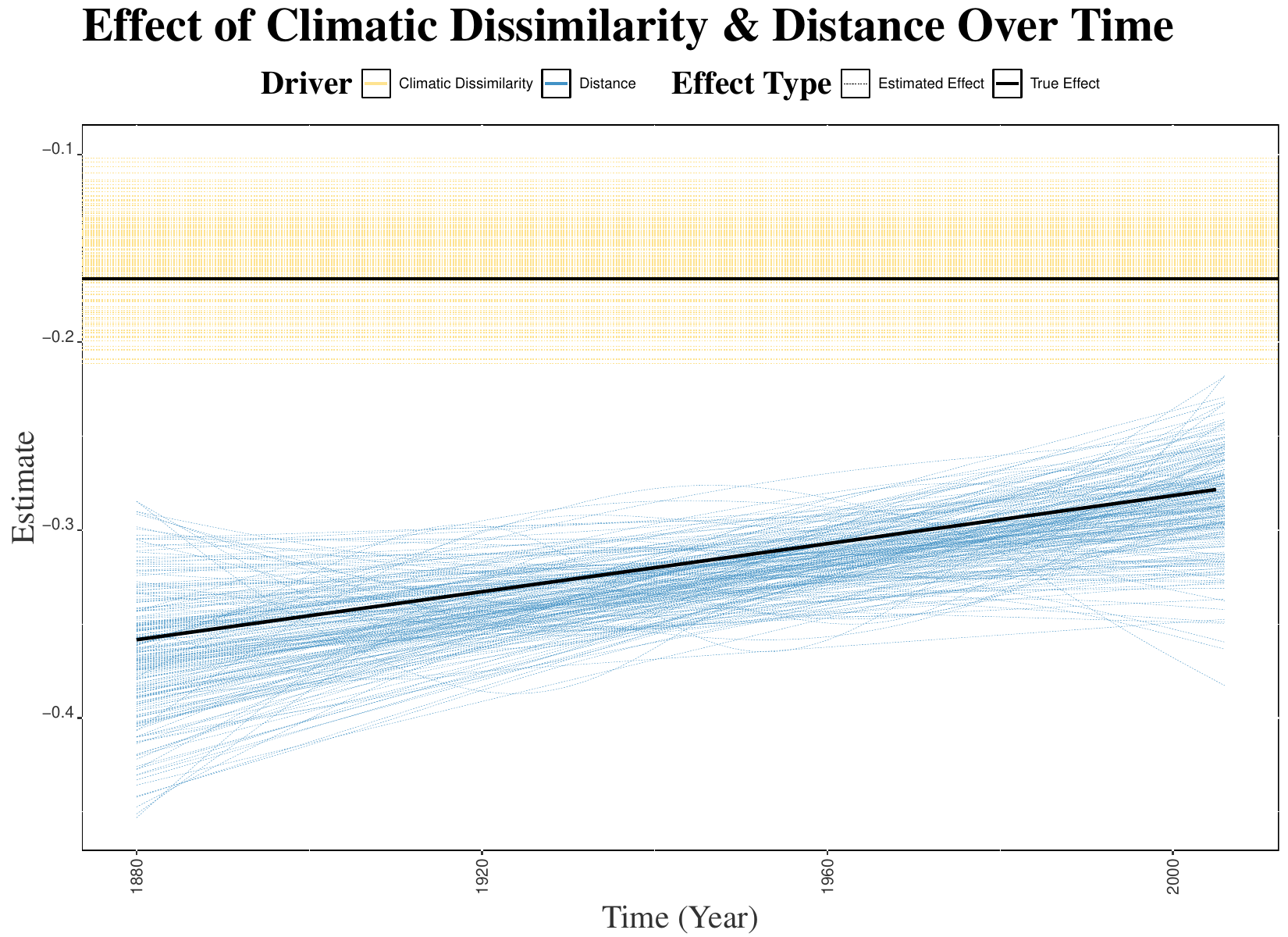}
		&
		\includegraphics[height=6.7cm]{ 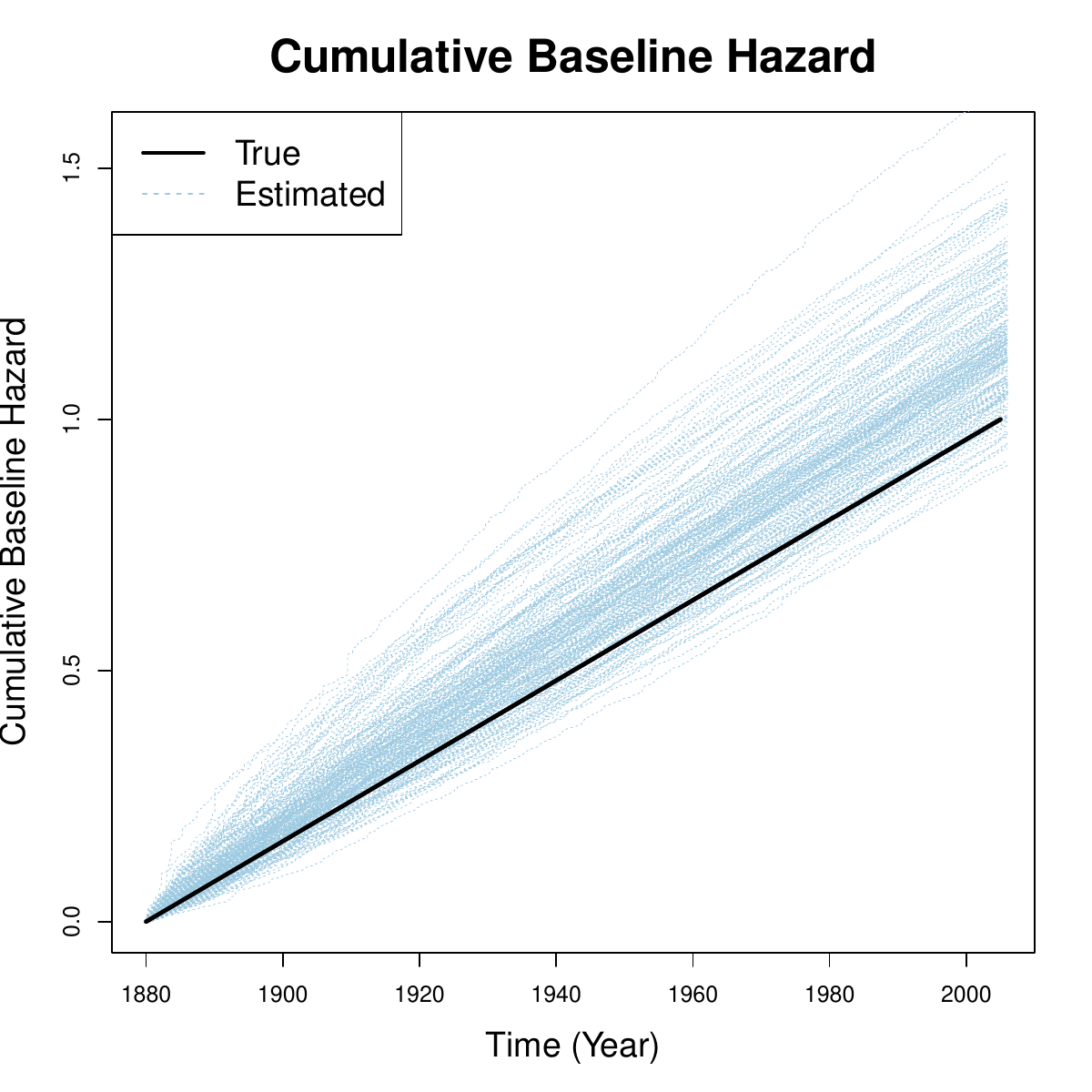}\\
		(a) & (b) \\
		\includegraphics[height=6.7cm]{ 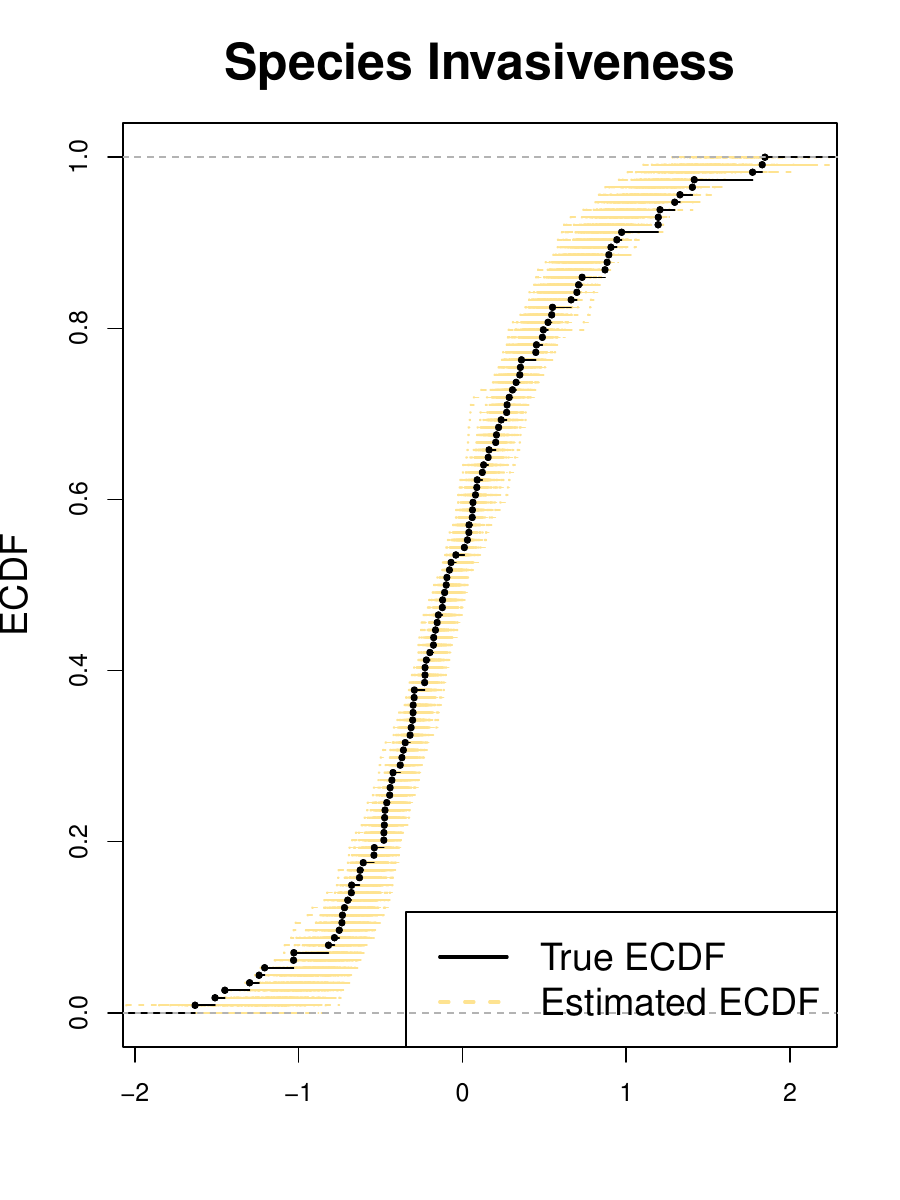}
		\includegraphics[height=6.7cm]{ 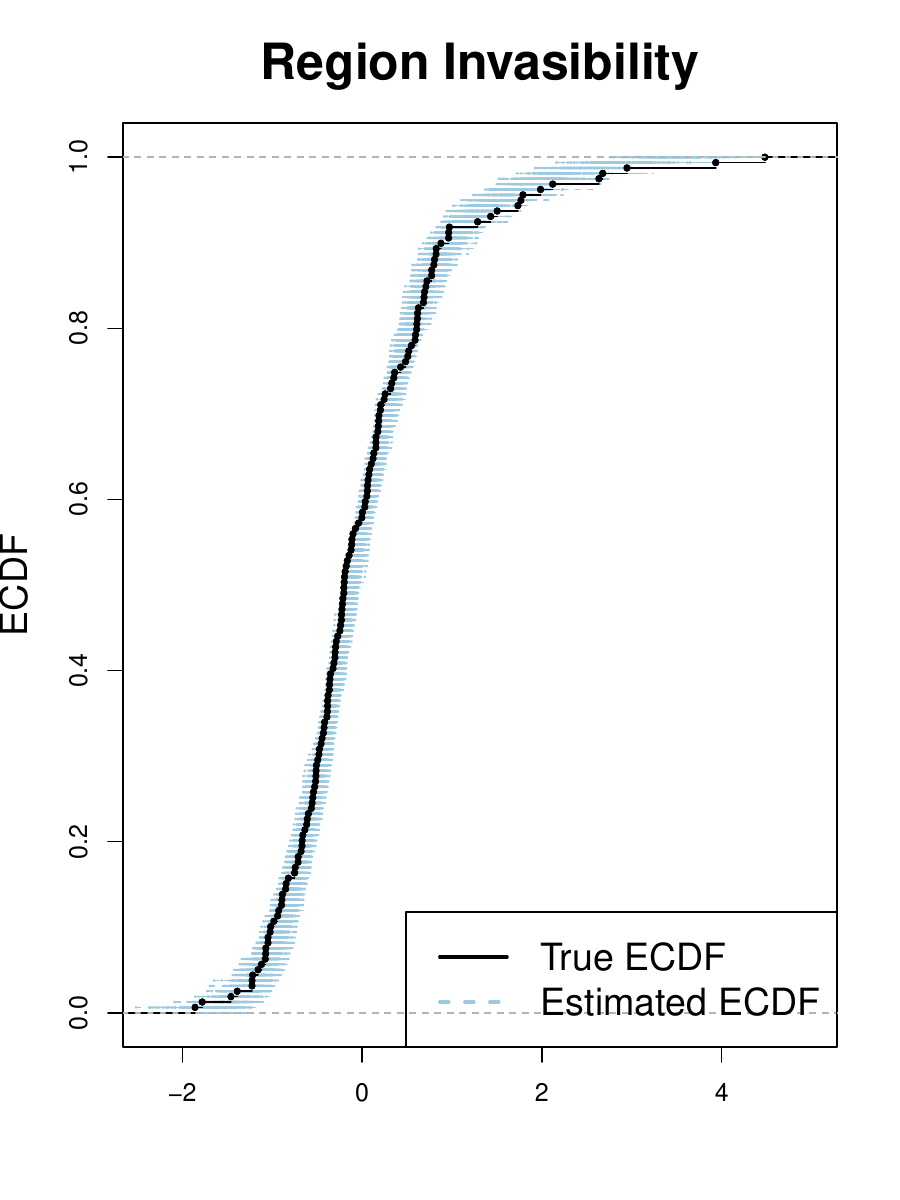}
		&
		\includegraphics[height=6.7cm]{ 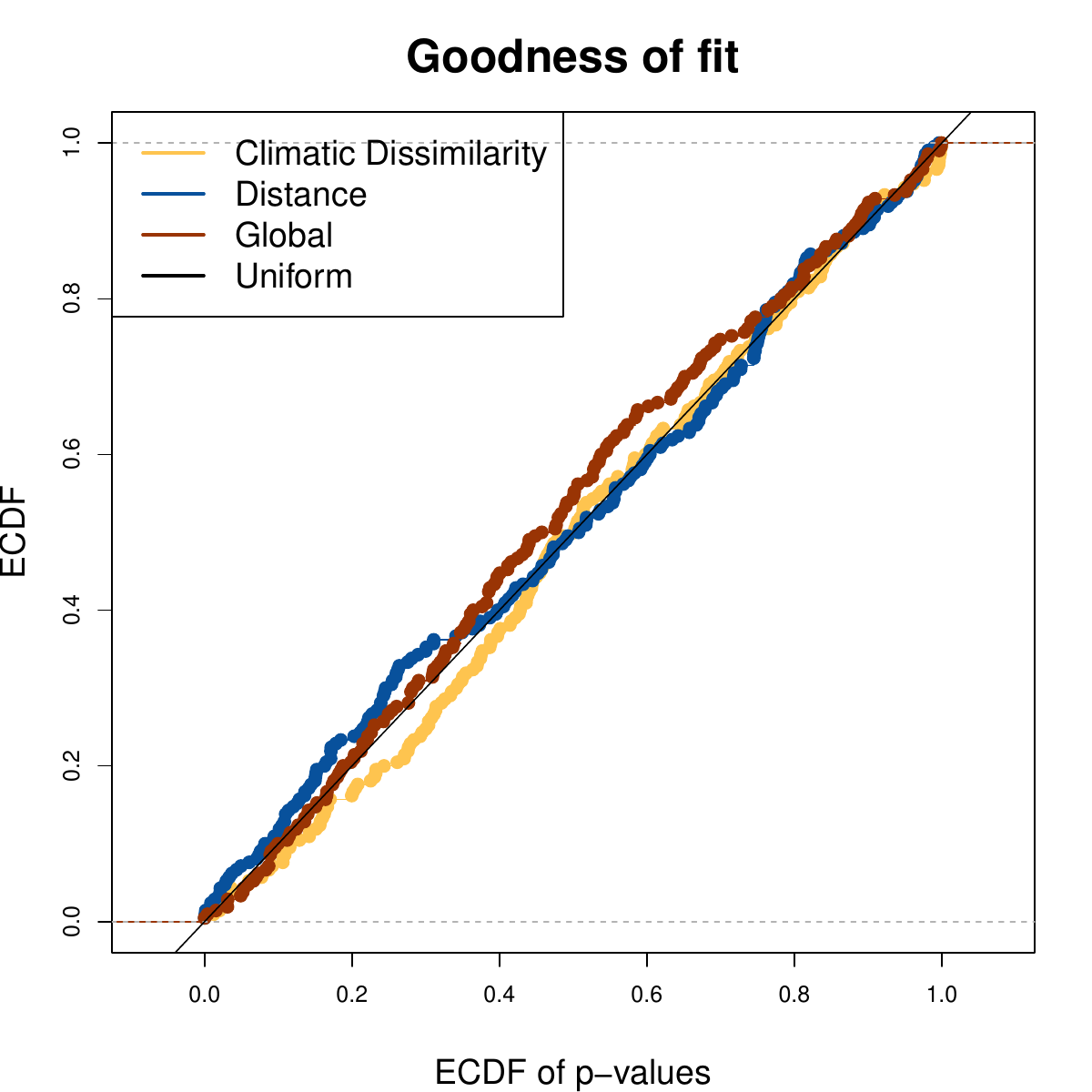}\\
		(c) & (d) \\
	\end{tabular}
	\caption[Simulation Study: Summary of the Results]{\label{fig:sim-results} \textbf{Simulation Study: Summary of the Results}. a) Comparison between the true coefficients (black solid) and the estimated coefficients for \textit{climatic dissimilarity} (yellow dashed) and \textit{distance} (blue dashed) on the simulated data. b) Comparison between the true (black solid) and the non-parametric estimates (blue dashed) of the cumulative baseline hazard. True baseline hazard is assumed to be constant and equal to $0.008$. c) Comparison between the true (black solid) and estimated random effects for species invasiveness (yellow) and region invasibility (blue). True values are represented by the conditional expectation of random effects fit on real data, while estimates correspond to the  $0$-dimensional spline estimates on the simulated data. Comparison has been performed in terms of empirical cumulative distribution function. d) Empirical distribution of the p-values resulting from testing \textit{climatic dissimilarity} (yellow), \textit{distance} (blue) and their \textit{global test} (brown). The empirical distributions are compared with the uniform cumulative distribution (black), which is the expected distribution of the p-values when the model is adequate.}
\end{figure}

\section{Simulation Study}\label{sec:simulation}

In order to show that the available data and model formulation are adequate to draw meaningful conclusions, we provide two simulation studies. First we would like to evaluate the inference techniques proposed in this paper for the type of empirical setting considered. Secondly, we want to evaluate the possible effect of model misspecification. As the recording of invasive species may not be uniform across the globe, we want to evaluate this potential recording bias on the parameter estimates in the model.   

\subsection{Validation of Introduced Techniques}

In this section, we aim to evaluate (i) the estimation of the mixed additive relational event model, (ii) the non-parametric technique for evaluating the baseline hazard function, and (iii) the strategy for assessing the goodness-of-fit \emph{within a data-setting similar to the one used in practice}. To achieve this, we first fit a basic model that includes \textit{distance}, \textit{climate dissimilarity} for insect invasions, obtaining estimates for the time-varying effect of the former, and a constant effect of the latter.  We also include random species and regions effects. We define a native range $\mathcal{NR}(1880)$ similar to the observed process and set the risk set in 1880 as its complement, $\mathcal{R}(1880) = \mathcal{NR}(1880)^c$. 

We then proceed to simulate complete IPs $210$ times for all insect species, using these parameter and random effect values over a time period from 1880 to 2005 via Gillespie-type algorithm \citep{gillespie1977exact}. Inter-arrival times are simulated from a continuous exponential distribution with piecewise-constant rates,
\begin{displaymath}
	\begin{aligned}
		\lambda_{sr}(t)  =& 0.008 \times  
		 \exp{\left[ \beta_{dt} \cdot dt_{sr}(t) + \beta_d(t) \cdot d_{sr}(t) + b_{s} + b_{r} \right]} \\
	\end{aligned}
\end{displaymath}
where $ \beta_{dt}, \beta_d(t), b_s$, and $b_r$ were estimated by fitting a mixed-effect additive REM on FR data involving insects. We obtain a number of invasion events ranging from $3847$ to $4399$.

For each simulated data set, we perform a NCC inference procedure, estimating the fixed parameters and random effects, the baseline hazard, and performing goodness-of-fit estimates. We sample a case-control data set, with the number of rows equal to the number of simulated events, a response column of $1$s and columns with covariate differences $\Delta \bm{x}$ and $\Delta \bm{z}$ for the event and randomly sampled non-event. Furthermore, time is discretised in a way simulated data can match actual data as closely as possible. We fit a additive mixed-effect logistic regression to obtain the parameter estimates $\hat{\beta}_{dt}, \hat{\beta}_d(t)$ and conditional expectations of the random effects $\hat{b}_s, \hat{b}_r$. Figure~\ref{fig:sim-results} a) and c) show that NCC estimation procedure achieves unbiased results. Estimates for cumulative baseline hazard in Figure~\ref{fig:sim-results} b) show that the method is adequately capturing the underlying trend. Finally, the goodness-of-fit plots in Figure~\ref{fig:sim-results} for \textit{climatic similarity}, \textit{distance}, and their global test illustrate that the proposed method adequately captures the model's goodness-of-fit. There is no evidence that the discretisation of time is negatively impacting the quality of the results.

\subsection{Recording Bias}

\begin{figure}[tb]
	\centering
	\includegraphics[height=7cm]{ 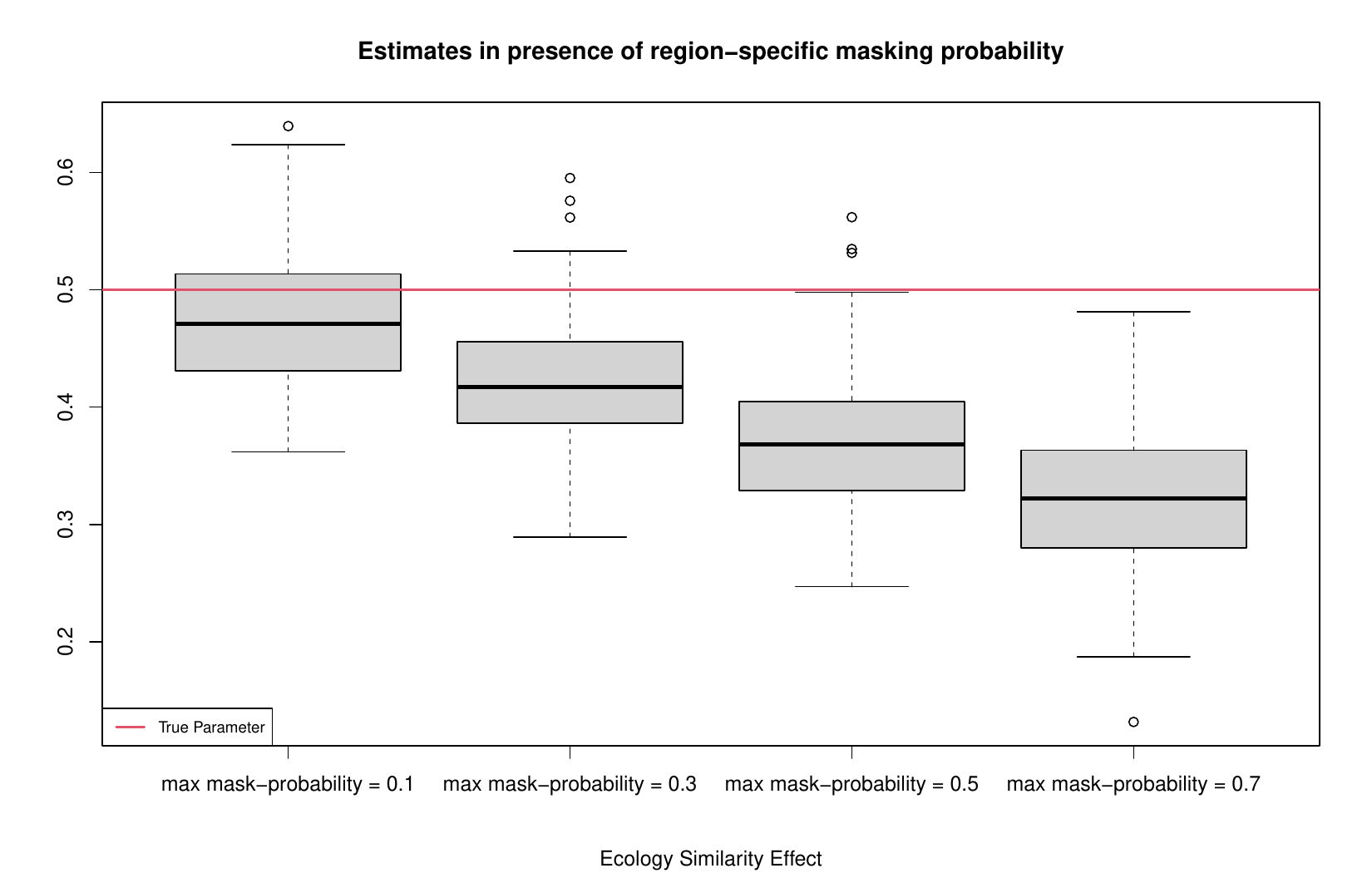}
	\caption[Simulation Study on Recording Bias]{\label{fig:recording-bias} \textbf{Simulation Study on Recording Bias}. Estimated coefficients from 100 experiment replications are compared to the true parameters used in data simulation. The maximum masking probability represents the upper limit of region-specific masking probabilities, ranging from 0 to this maximum value. While increasing the maximum masking probability shrinks the estimates towards zero, the sign of the effect remains correctly identified.}
\end{figure}

The second simulation study evaluates the behaviour of our estimation procedure in the presence of recording bias. As highlighted by \citet{bonnamour2021insect}, variation in scientific recording efforts can influence our understanding of alien species diffusion dynamics. To assess this effect, we simulate species invasion as a function of an endogenous driver, e.g., ecological similarity between the destination region and the most recent source region where the species was detected. To account for differences in sampling effort across regions, we introduce a masking probability for each region and randomly remove a portion of the simulated invasions accordingly. Figure~\ref{fig:recording-bias} presents the estimated coefficients from 100 experimental replications against the true parameter used for data generation. Even under strong masking probabilities, our methodology correctly identifies the sign of the coefficients. In the empirical application, we mainly focus on the interpretation of the sign and trend of the effects, rather than their absolute magnitudes. We also observe that increasing masking strength shrinks coefficient estimates toward zero. Nevertheless, even when the majority of the records from certain regions is missing, the direction of the effect remains correctly identified and does not overlap with zero for any instance.

\section{Analysis of Plant and Insect Co-invasions}\label{sec:application}


The main substantial goal of this paper is to understand which forces shape the diffusion of alien species across the globe. This is a complex process with many feedback loops that have typically been ignored in previous studies. By properly accounting for the endogenous and temporal nature of the process, our model captures the complexity of the joint insect and plant invasion process. Plants and insects have diverse and complex relationships. 

Beneficial interactions between plants and insects are typically referred to as mutualism \citep{bronstein2006evolution}.
However, not all interactions between plants and insects are mutually beneficial. 
Many insects are herbivores and feed on plant parts such as leaves, stems, flowers, or fruits. \emph{Herbivory} can cause direct physical damage to plants, leading to reduced growth, impaired photosynthesis, or even death. Some insects, such as spongy moth larvae and tent caterpillars, specialize in defoliating plants by consuming their leaves \citep{hemming1995intraspecific}, whereas other insects, such as aphids and mealybugs, feed on plant sap by inserting their mouthparts into plant tissues and extracting nutrients \citep{branco2023sap}. Some insects consume plant seeds, affecting the plant's reproductive success. Another type of non-mutualistic interaction is \emph{gall formation}. Certain insects induce the formation of abnormal growths, called galls, on plants, manipulating its physiology to create a protective structure that provides the insect with food and shelter \citep{takeda2021recent}. Finally, insects can act as vectors of \emph{disease transmission}. They may pick up pathogens from infected plants, and transmit them to healthy plants while feeding or through physical contact. Aphids play a central role in virus transmission, that may lead to plant diseases, manifesting, for example, with yellowing \citep{sankarganesh2020chapter}.
Differential mutualisms between plants and insects seem to have a role in alien species invasions \citep{prior2015mutualism,simberloff1999positive}. The term invasional meltdown has been introduced for how non-indigenous species may collaborate to increase the likelihood of a successful invasion. For instance, non-native plants and insects may cause variations in their new habitats enhancing the opportunity for other non-indigeneous species to get established in these areas.

\begin{figure}[t]
	\centering
	\includegraphics[width=0.7\linewidth]{ 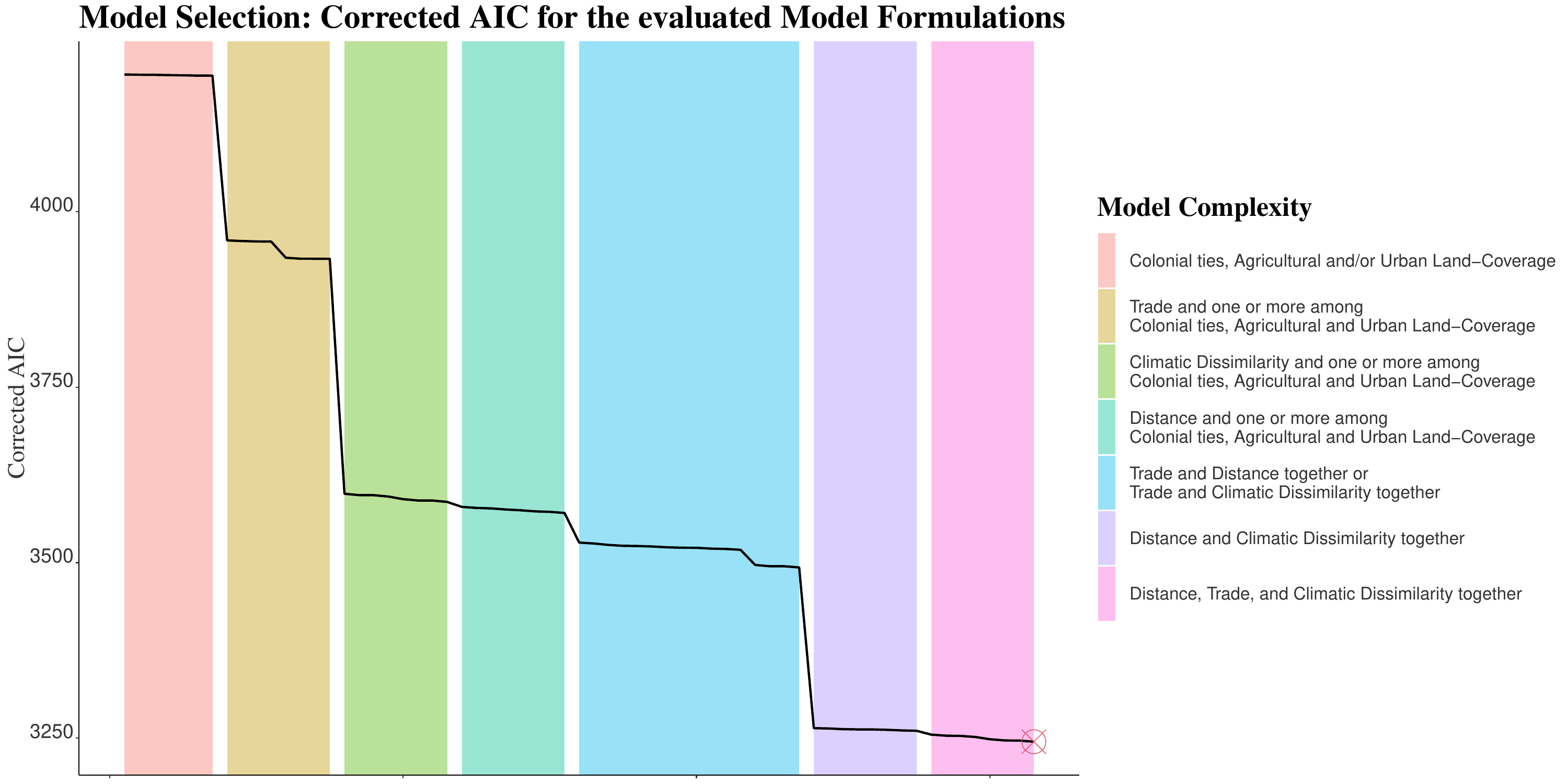}
	\caption[Model Selection]{\label{fig:AIC-values} \textbf{Model Selection}. Values of AIC for the examined model formulations. We outline that, whenever included, \textit{distance}, \textit{trade}, and \textit{agricultural land-coverage} are supposed to have a time-varying impact while \textit{climatic dissimilarity}, \textit{urban land-coverage}, and \textit{colonial ties} a fixed effect. This choice comes from the previous studies on the topic \citep{Juozaitiene2022}. On the other side, all the considered models include random effects for species invasiveness, region invasibility and species co-invasion. The best model in terms of corrected AIC \citep{woodAIC} includes \textit{distance}, \textit{trade}, \textit{colonial ties} and \textit{climatic dissimilarity} and is outlined in the plot with a red crossed symbol. According to the covariates included in the compared models, we can distinguish seven groups of model formulations.}
\end{figure}

In the species invasion event graph, the species nodes $ \mathcal{S} $ can be separated into two distinct types, i.e., the ensemble of insects $ \mathcal{S}_{\mbox{\scriptsize ins}} $ and the set of vascular plants $ \mathcal{S}_{\mbox{\scriptsize plt}}$. The receiver nodes in the species invasion event graph are the countries or geographically defined regions. As a species invades a region, a time-stamped directed edge arises from the set of species $\mathcal{S}$ towards one of the nodes in the collection of regions $ \mathcal{C} $. Following \citet{borgan1997risk}, we stratify our model using the taxonomy $ a $ (vascular plants vs. insects) of the species involved, as in Equation $ \eqref{3} $.

One of the aims is to study alien plant and insect \emph{co-invasions}. In addition to the drivers reported in Table~\ref{tab:covariates}, we introduce the three random effects. Besides the two main effects, species invasiveness and region popularity, we define a species interaction effect, $b_{ss'}$, which captures how much the presence of last species $s'$ in a certain region stimulates or impedes the invasion by species $s$ into that region. These two entries can assume the categories \textit{Rare interaction} and \textit{Novelty} as well. The former is used when the species-last species interaction appears only once among those recorded for events and non-events; the latter is taken into account if no other species are detected in the related country before the considered time. We point out that species and last species, in this context, may be members of separate taxonomies, allowing us to study possible symbiotic relationships between insects and vascular plants. The species interaction effect further underlines that the conditional independence assumption does not imply independence between events in this model, as past events can influence current ones. However, these relationships are assumed to be fully explainable through the history of previous occurrences.

\begin{figure}[b]
	\begin{tabular}{cc}
		\includegraphics[width=0.48\linewidth]{ 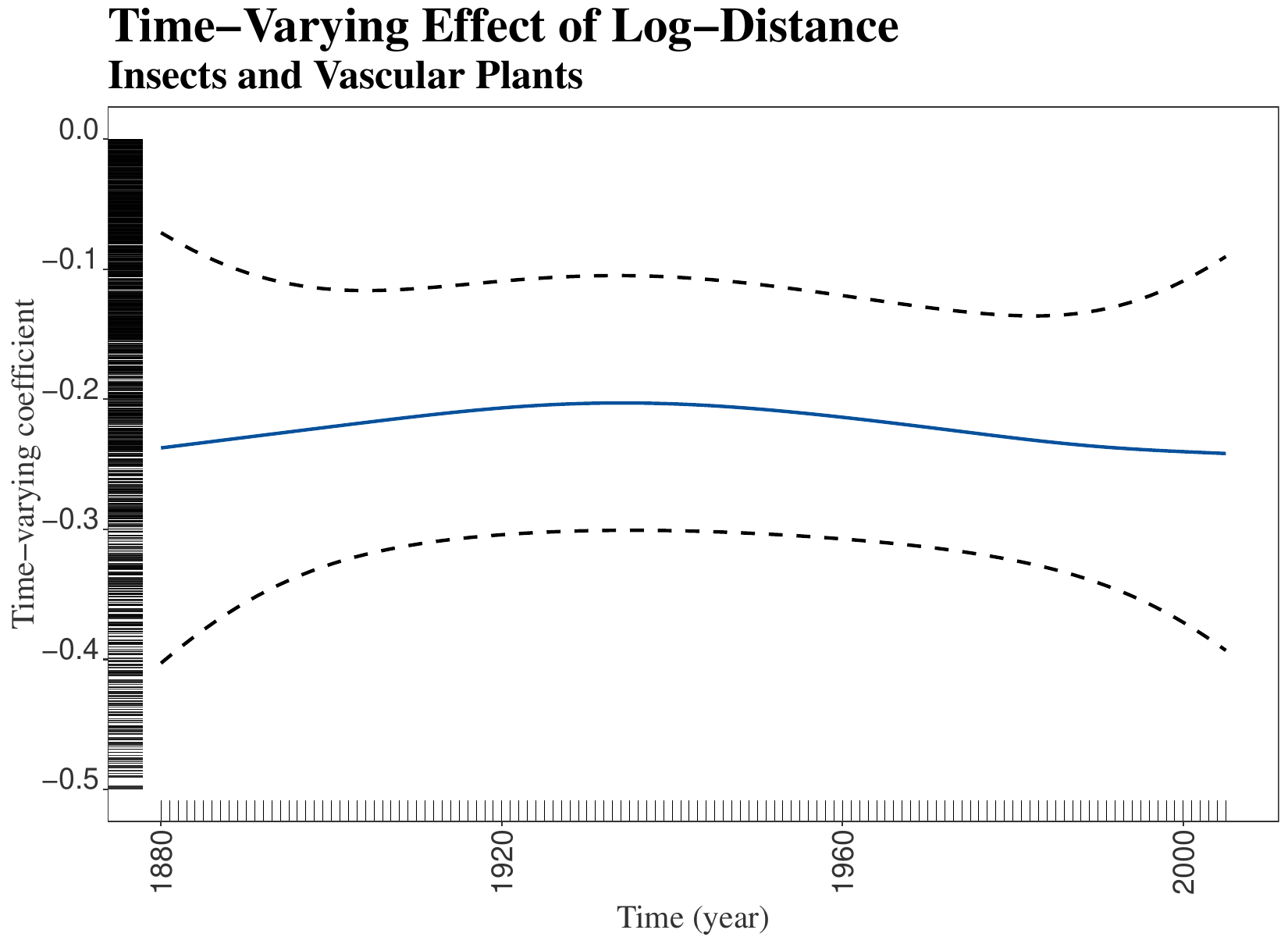}
		&
		\includegraphics[width=0.48\linewidth]{ 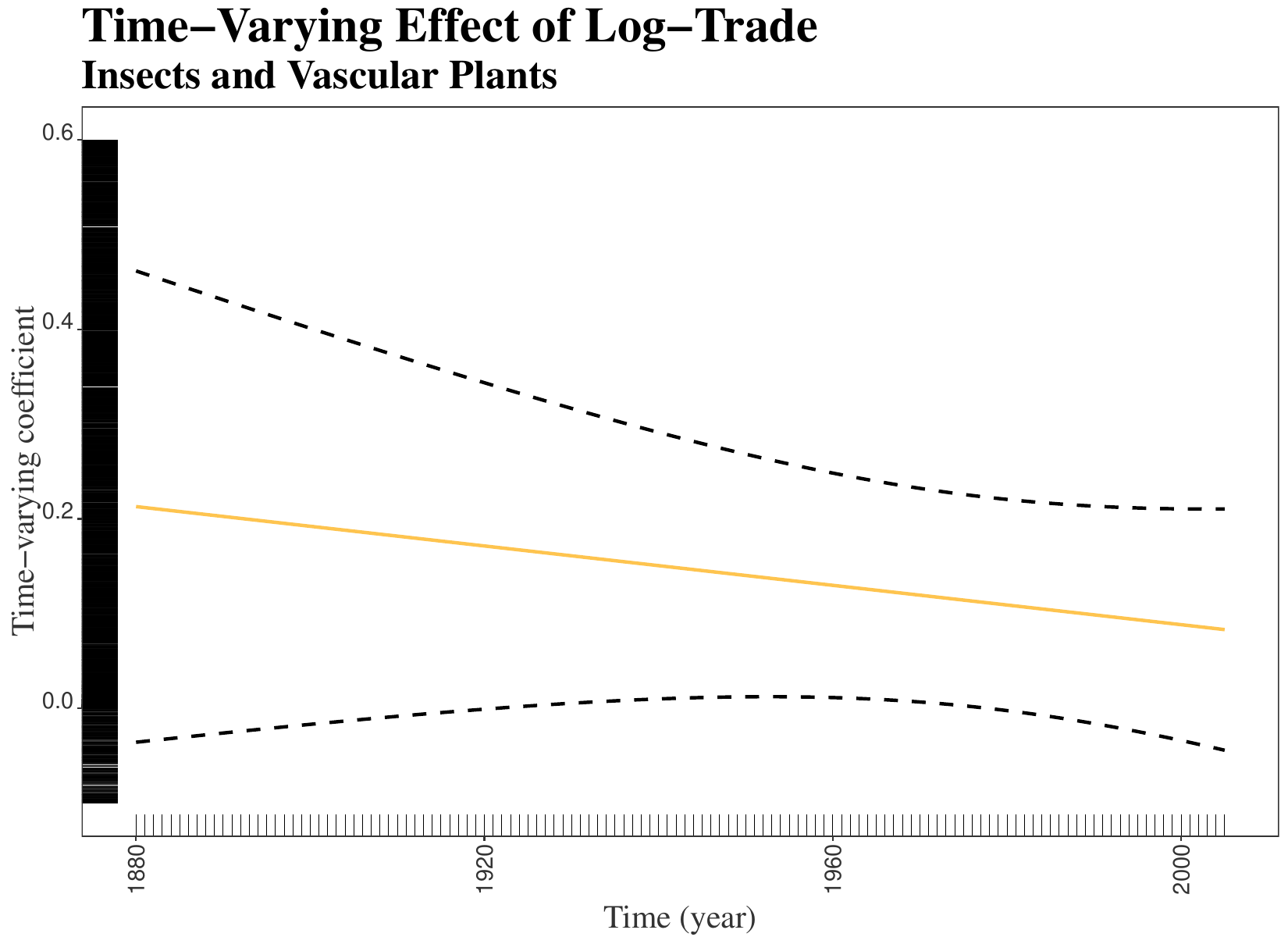} \\
		(a) & (b)
	\end{tabular}
	\caption[Time-varying estimates]{\label{fig:tve} \textbf{Time-varying estimates}. Time-varying estimated coefficients for a) \textit{distance} and b) \textit{trade} (undashed lines) with the related posterior confidence intervals (dashed lines).}
\end{figure}

Model selection has been conducted by evaluating the corrected version of AIC. This correction, relying on the adjustment for the degrees of freedom, avoids chiefly selecting the simplest or the most complex model. Figure $ \ref{fig:AIC-values} $ shows the values of AIC for the $ 63 $ evaluated model formulations, including different subsets of the covariates shown in Table~\ref{tab:covariates}. Following \citet{Juozaitiene2022}, whenever included, \textit{distance}, \textit{trade} and \textit{agricultural land-coverage} have a time-varying effect, while \textit{climatic dissimilarity}, \textit{urban land-coverage} and
\textit{colonial ties} remain constant over time. The best model according to corrected AIC includes \textit{distance}, \textit{trade}, \textit{colonial ties} and \textit{difference in temperature}.  Land-cover characteristics seem to have no impact on the dynamics of invasions. Both the fixed coefficients related to the climatic conditions and to the spread of the colonialism turn out to be negative. The former result is quite intuitive: species tend to invade regions that have similar climatic characteristics as the regions they have invaded before. The latter instead is somewhat counter-intuitive, but it could be a residual effect as trade has a strong positive effect. Perhaps surprisingly, as shown in Figure~\ref{fig:tve}b, the effect of trade seems to be diminishing over the last century. \cite{Juozaitiene2022} suggest that this may be the result from the fact that international trade is now more focused on products rather than on raw materials. Additionally, stricter international transportation regulations may have led to a decrease of the effect of trade on species invasions. The time-varying effect related to \textit{distance} is negative as seen in Figure~\ref{fig:tve}a, and remains relatively constant between 1880 and 2005. This fact confirms how unusual long-distance invasion occurrences are. 

\begin{figure}[t!]
	\centering
	\includegraphics[width=0.7\linewidth]{ 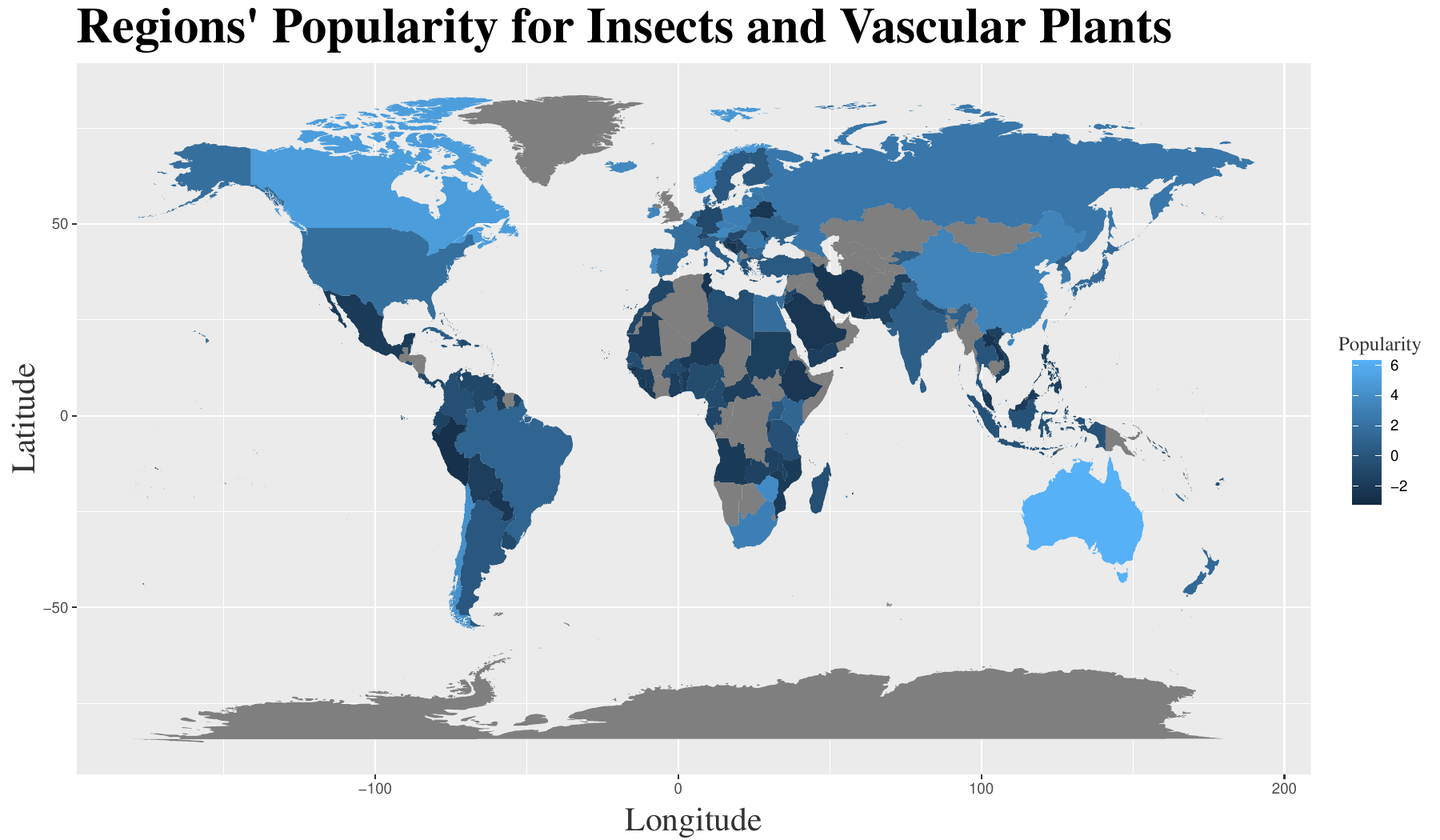}
	\caption[Regions' Invasibility]{\label{fig:insects-map} \textbf{Regions' invasibility in terms of estimated random effects}. Lightest-blue areas are those that the model identifies as most popular, such as Australia and Canada. On the other hand, darker locations are those that lead to a decrease in the rate of occurrence of alien species invasions (Peru and Saudi Arabia are some instances).}
\end{figure}

\begin{figure}[tb]
	\centering
	\includegraphics[width=0.7\linewidth]{ 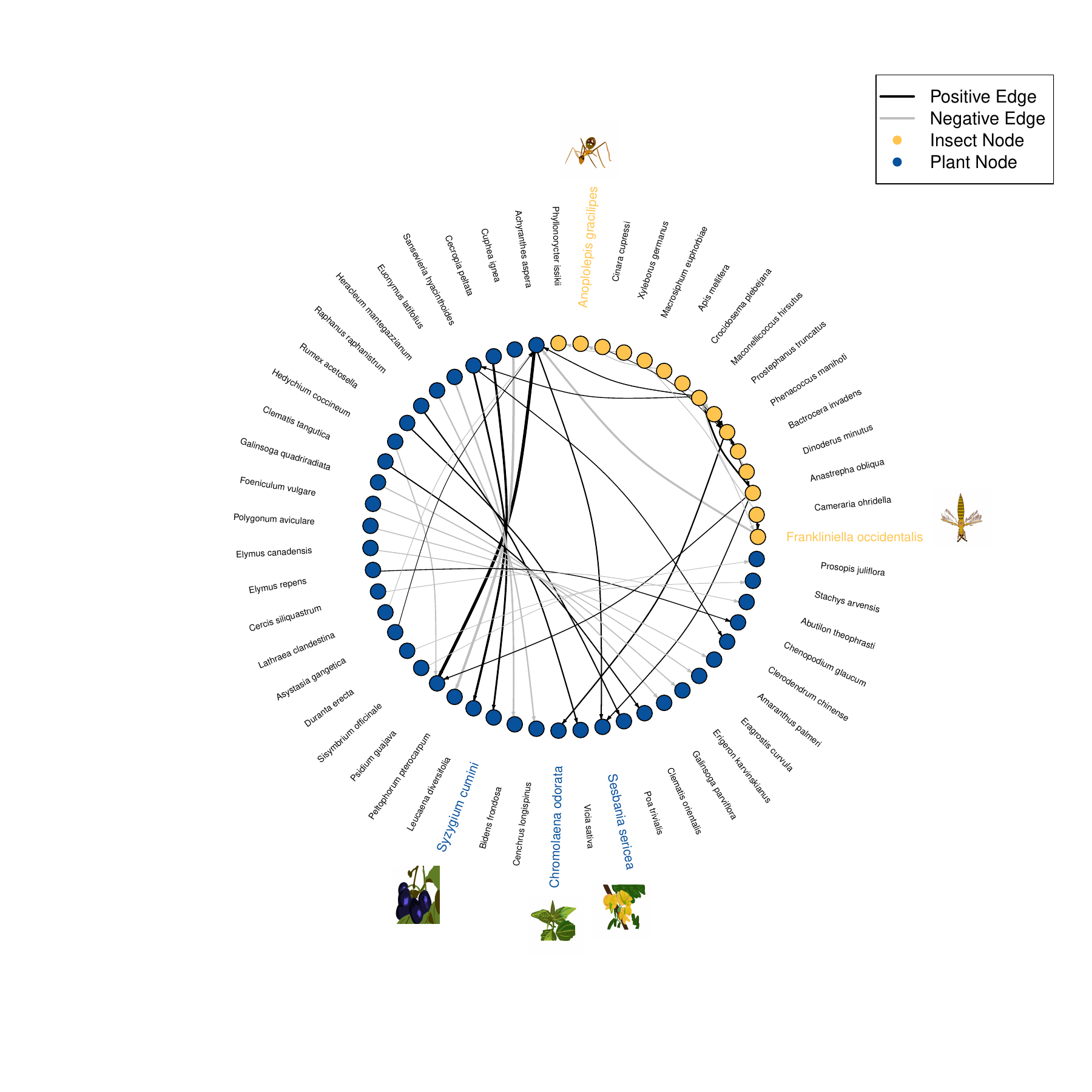}
	\caption[Associations between Species in terms of Co-invasion]{\label{fig:species-coinvasion} \textbf{Strongest associations between species in terms of co-invasion}. We report here the values that are equal or larger, in absolute value, than the logarithm of $1.5$. Taking into account all the other features, the presence of the linked species-last species co-invasion lead to increasing or decreasing the risk of invasion at least of $ 50 \% $. The taxonomy determines the colour of the nodes (blue for plants and yellow for insects). When the estimated random coefficient is positive, the link is shown in black; a positive link indicates a higher probability of the sender node following the receiver node, meaning the risk for the sender species of entering a nation that has just been invaded by the recipient species in the plot is increased. The thicker the edge, the larger (in absolute value) the estimate for the corresponding random effect. The figures represent the nodes with the largest values in terms of species invasiveness.}
\end{figure}

Once the fixed and the time-varying effects of the covariates are properly accounted for, one can consider the conditional expectation of the random effects related to the insect and plant invasiveness. \emph{Frankliniella occidentalis}, the most invasive insect found, has a long history of foreign species incursions dating back to the $ 1970$s. It is widely known for its pest qualities and ability to cause significant plant harm \citep{Frankliniella}. Another largely invasive species is \emph{Anoplolepis gracilipes}, a very small ant, which negatively impacts native ecosystems, particularly forests \citep{lee2022biology}. Among the most invasive plants represented in Figure~\ref{fig:species-coinvasion} is \textit{Chromolaena odorata}. While its negative effects in South Africa have been studied extensively, leading to its classification as a controlled species, it has also been noted for its beneficial effects in central Africa \citep{goodall1996review}. The random effects of insect and plant invasiveness have, respectively, standard deviations of 1.86 and 0.31, suggesting that the heterogeneity on the insect side is markedly higher than on the plant side. Figure~\ref{fig:insects-map} depicts the regions’ invasibility by means of their conditional random effects. Australia and Canada have the highest invasibility, but also South Africa, United States, and New Zealand have high values, meaning that the probability of alien species invasions in these regions is larger than in other areas.

Figure~\ref{fig:species-coinvasion} also illustrates the strongest co-invasion relationships between the species. Particularly, it shows whether the presence of a species affects the rate of an invasion event by another species. Considering interactions between species that pertain to different taxa, we see a positive effect for \emph{Phenacoccus manihoti}, commonly known as the cassava mealybug,  when \emph{Chromolaena odorata}, or Siam weed, has reached the country. \citet{calatayud} studied  the variations in the dispersion dyamics of  \emph{P. manihoti} in relation to other factors in Brazzaville (Congo), where \emph{C. odorata} is the primary plant species.  The opposite tendency, i.e., a negative interaction effect, is found for \emph{Frankliniella occidentalis} (western flower thrip) invading a region that has already been reached by \emph{Achyranthes aspera} (chaff-flower). We can find in the literature two instances where the aforementioned weed seem related to viruses whose transmission \emph{F. occidentalis} may be involved. Particularly, \emph{A. aspera} is specifically mentioned by \citet{kumar2008emergence} as being a part of the \emph{Tobacco Streak Virus}' native range. Because of the harm caused by this virus, one potential response may be the development of a resistance, which has actually been observed in a plant that is resistant to \emph{F. occidentalis}.
Additionally, the analysis of the plants in the tomato production area in Kenya in \citet{macharia2016weed} revealed the presence of \emph{A. aspera}. However, it is unclear how it might act as a host for the \emph{Tomato spotted wilt virus}, for which \emph{F. occidentalis} is a renowned vector.

\begin{figure}[b]
	\begin{tabular}{cc}
		\includegraphics[width=0.5\linewidth]{ 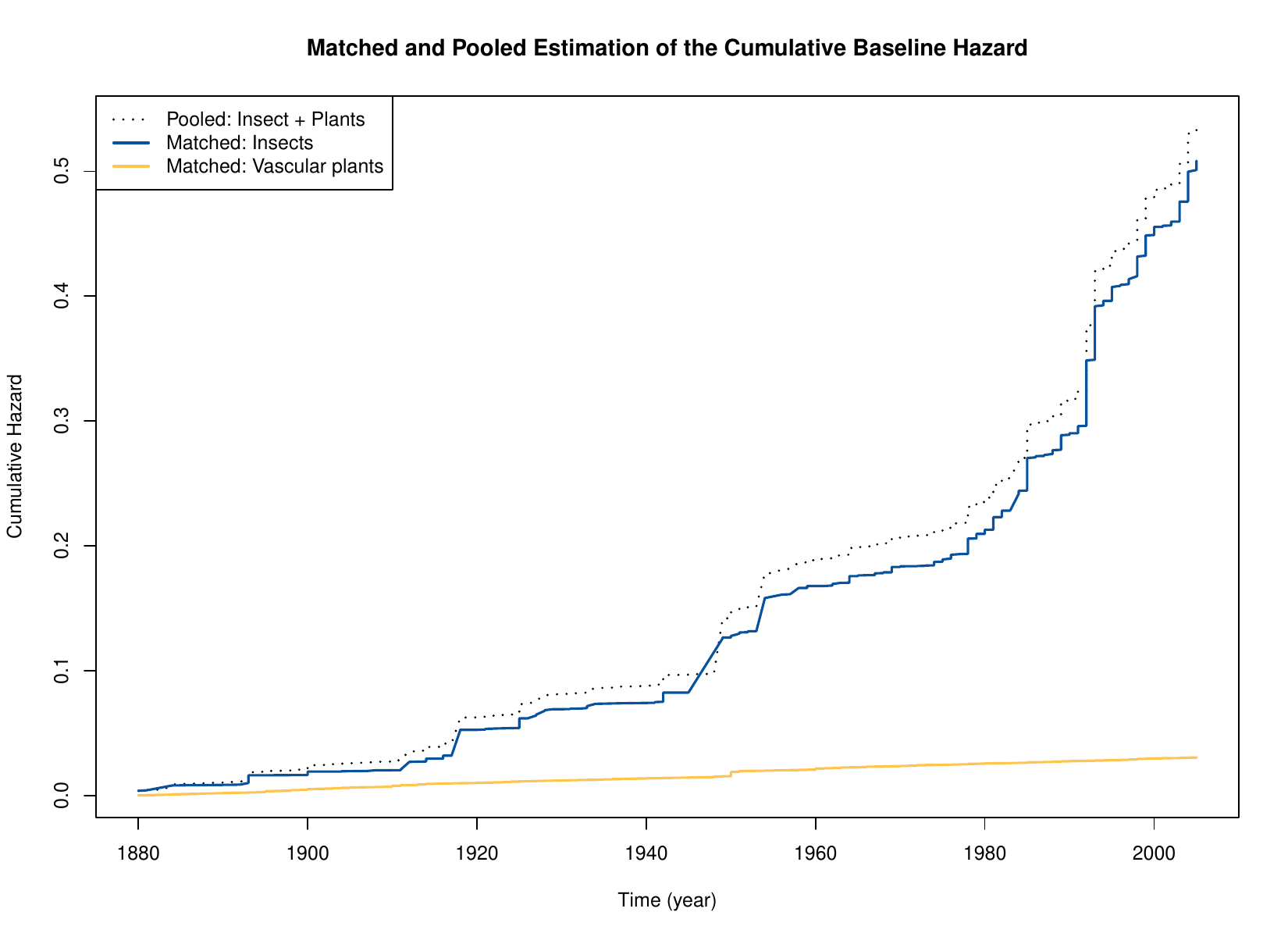}
		&
		\includegraphics[width=0.5\linewidth]{ 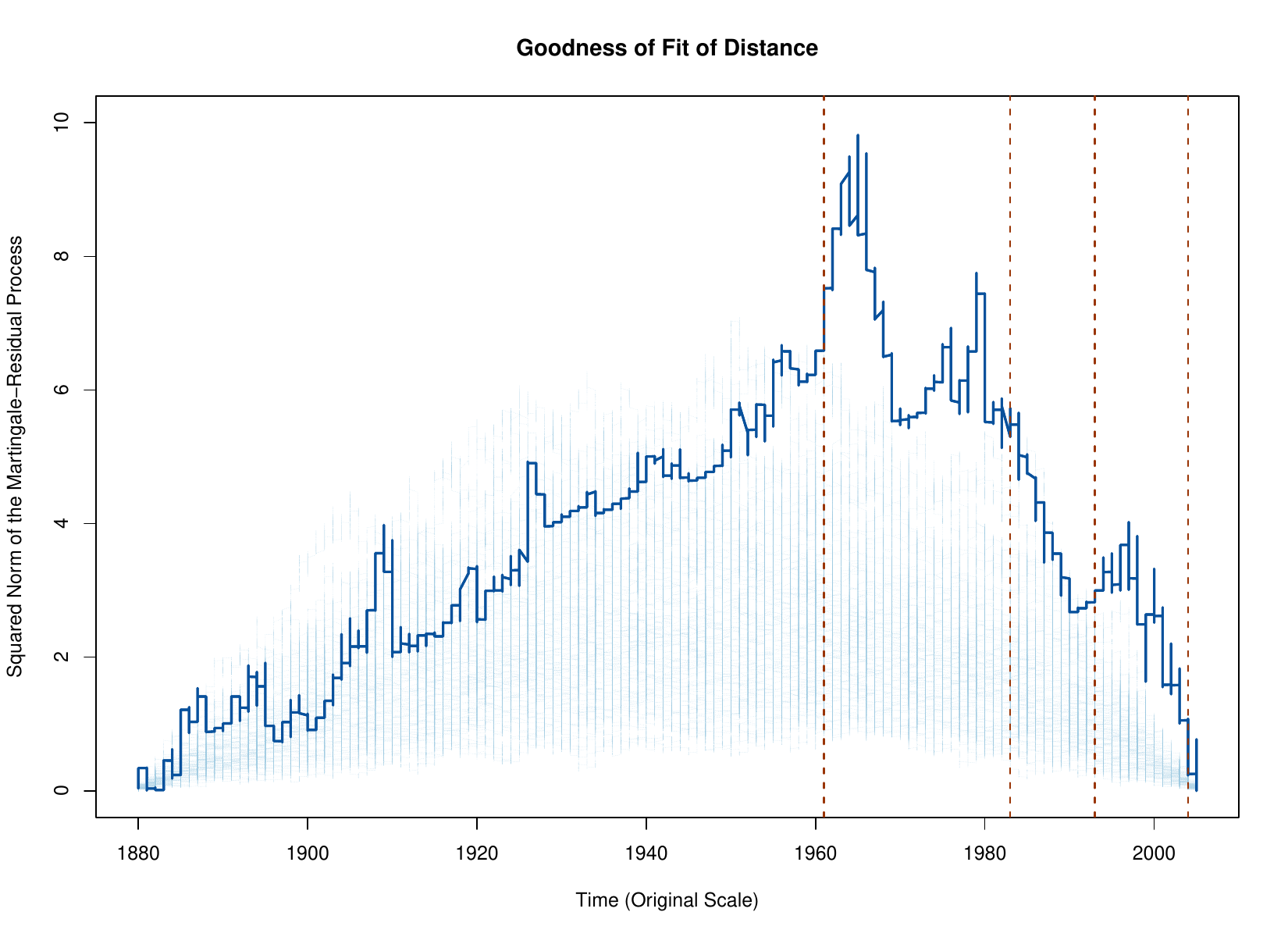}
	\\
	(a) & (b)
	\end{tabular}
	\caption[Estimated Cumulative Baseline Hazard and Goodness of Fit of Distance Covariate]{\label{fig:baseline} a) \textbf{Matched and Pooled Estimates of the Cumulative Baseline Hazard}. Following \citet{borgan1997risk}, we may provide a taxa-specific estimate of the baseline (blue for insects, yellow for plants, solid); as an alternative, we may rely on a pooled estimate (dotted) that takes into account the fact that the sampled control belongs to the same group as the observed case. b) \textbf{Goodness of Fit of Distance Covariate}. Graphical comparison between the observed squared norm of the martingale-residual process against $200$ simulated squared norm of Brownian Bridges. We plot the time in the original scale in order to understand which section of the time-window may lead to misspecification. It is interesting to see that the Bridge overcome the cloud created by the theoretical curves in the time range between $1961$ and $1983$ and in the time window between $1993$ and $2004$. We hypothesize that the impact of distance may have evolved in recent times, suggesting that more complex, non-linear, effects of distance, could improve the model fit.}
\end{figure}

Once all the features have been taken into account, Figure~\ref{fig:baseline} represents the estimated cumulative baseline hazard according to \eqref{11}. The shape of the curve for insects suggests that the rate of invasions for this taxonomy has been growing strongly in recent years, in a way that the above covariates are unable to account for. Yet, for vascular plants, the existence of a roughly linear baseline hazard may be interpreted that most of the important covariates for plants are accounted for in the selected model. This result is in line with similar results reported in the literature. \citet{bonnamour2021insect} report that the invasive behaviour of insects in the past decades has been stronger than that of plants. Particularly, they claim this is due to the fact that insects took more advantage from the availability of fast transportation tools rather than plants. Moreover, insects are able to survive for longer across long journeys. \citet{walliser2013attracting}, furthermore, argued that due to their smaller dimensions and inherent mobility insects may also be harder to spot and control than plants.

\begin{table}[b]
	\centering
	\begin{tabular}{rlrr}  
		\hline & Covariate & p-value  & Dimension $q$ \\  
		\hline & \textit{distance} & 0.0002 & 10 \\  
		& \textit{trade} & 0.580 & 10 \\   
		& \textit{colonial ties} & 0.404 & 1 \\   
		& \textit{climatic dissimilarity} & 0.480 & 1 \\   
		\hline
		& Omnibus test & 0.468\\
		\hline
	\end{tabular}
	\caption[Goodness of Fit]{\label{tab:p-val} Evaluation of the goodness of fit for the each of the covariates included in the selected model. The table reports the p-values related to the KS test reported in Equation \eqref{15} ($q=1$) and \eqref{eq-KS} ($q=10$)}. 
\end{table}

Finally, we aim to understand if the included covariates correctly incorporate and describe the dynamics driving the alien species invasions. For this we will use the goodness-of-fit technique presented above. We use the KS test in \eqref{15} to test the fixed linear effects Colonial Ties and Climatic Dissimilarity. Instead, for Distance and Trade, we employ the KS test in \eqref{eq-KS} with $q=10$, as these effects are modelled via $10$-dimensional non-linear functions of time. Table~\ref{tab:p-val} reports the p-values of the associated tests. 
Importantly, the global test \eqref{eq:omnibus_test} is not rejected, with a $p$-value equal to $0.468$. In principle, this means that overall the model is adequate and no further tests have to be performed. If, for argument's sake, we would check the individual tests, we see that all covariates except \emph{distance} are adequately incorporated in the model. We hypothesize that the effect of distance might have changed in more recent times, as a result of faster and more varied modes of transport. Although we allow the effect of distance to change over time, i.e., $\beta_d(t) \cdot d_{sr}(t)$, this may not have been sufficient and that the effect of distance change both over time and over distance itself in a non-linear way, i.e., $f\left( t, d_{sr}(t) \right)$. As mentioned above, the $p$-value is obtained empirically, by emulating the theoretical behaviour of the multivariate Brownian Bridge. We can thus compare graphically the observed $\lVert \bm{M}[\hat{\bm{\gamma}}, u|\mathcal{E}] \rVert^2, u\in [0,1]$ against the squared norms of Brownian Bridges $\sup_{u\in[0,1]}{\lVert  \bm{Z}^0(u) \rVert^2}$, as shown in Figure~\ref{fig:baseline}b.

\section{Discussion}\label{sec:conclusions}

The mixed additive relational event model is a rich explanatory and quasi-generative model of the dynamics driving alien species invasions. It is able to integrate and account for various ecological characteristics and socio-economical features. Although not directly comparable, our findings align with various, more qualitative studies in the literature. For instance, \citet{chapman2017global} emphasize the role of trade networks in explaining the large-scale distribution of non-native species, particularly when weighted by proximity and climatic similarity. Similarly, \citet{mwebaze2023role} highlight the interaction between trade and climatic conditions, arguing that trade between countries with similar climates poses a greater invasion risk. These results align with our findings, which identify a positive effect of international \textit{trade}, measured in a way that incorporates its underlying endogenous network structure. The negative effect of \textit{distance} suggests that species tend to diffuse over short distances, likely due to a higher probability of survival compared to invasions over larger distances. Moreover, the long-distance dispersal of invasive species is increasingly driven by human-mediated dispersal on a global scale \citep{zhang2023comparison}. Therefore, the negative impact of \textit{distance} may be partly explained by the implementation of legal instruments designed to mitigate invasion risks \citep{turbelin2017mapping}. Greater distances are also often associated with larger climatic differences, making it more difficult for species to establish in distant regions \citep{morlon2008general,buckley2008linking}. This is further supported by the negative effect of \textit{climatic dissimilarity}, which indicates that species tend to establish in regions with similar climatic conditions, where survival and adaptation  are more favorable \citep{seebens2015global,Juozaitiene2022}. This hypothesis is also reinforced by \citet{liu2020most}, who examined ecological niche conservation --- a widely accepted concept, albeit with some variability across taxa.

When analyzing first invasion records, we can track the potential occurrence of events at each time point using the risk set. The definition of the risk set at the beginning of the observational period is based on knowledge of species' native range  \citep{Juozaitiene2022}. Elements in the native range cannot be invasion events, by definition. The native range is also used, implicitly, compute several endogenous covariates that depend on knowledge of the regions where a species has already been previously recorded. These covariates, which are central to relational event modeling \citep{bianchi2024relational},  incorporate dynamic drivers that go beyond aggregated statistics, usually considered in this framework, such as the distance to the nearest invaded region. As highlighted by \citet{chapman2017global}, accounting for a species’ source region, rather than just its destination, is crucial.

Additionally, we have considered endogenous statistics that do not only rely on the source region, but also on other regions where the species might have been present before. For this reason, we argue that our method is able to address the so-called \textit{bridgehead} effect  \citep{bonnamour2023historical}. This is a region, where a species has been able to establish, and that serves as a source for new invasive species \citep{bertelsmeier2018recurrent}. Our model is able to incorporate explanatory variables that are computed not only at a species-region level, but also at a temporal level. Furthermore, their effects are also allowed to vary over time. Accounting for drivers whose influence changes over time is particularly relevant in ecology, especially when the observational period spans more than a century, as in our case. As \cite{bonnamour2021insect} highlight, the context of globalization has evolved significantly over this period, experiencing two major waves of globalization, leading to substantial changes in transportation systems and, consequently, in trade openness, both of which play a critical role in non-native species introductions. By modeling the effects as varying over time, we confirmed the findings of \citet{Juozaitiene2022}, which identified a decreasing impact of trade over time on the hazard of alien species first records. This trend has been attributed, following \citet{luppold1988hardwood} and \citet{mayer2003dynamic}, to trade shifts, including changes in commodity structures and the nature of traded goods, leading to a reduction in products that facilitate the spread of alien species.

Our model formulation integrates random effects to account for the heterogeneity of actors involved in both past and present events, as well as to model species co-invasions. In our approach, we include both monadic and dyadic random effects. Through this analysis, we identified \textit{Frankliniella occidentalis} as the most invasive insect, a finding supported by the literature \citep{Juozaitiene2022}. This is also recognized as a key pest from an economical point of view \citep{yang2015temporal}. Furthermore, species invasion random effects explicitly enables to account for co-invasion tendencies. This is particularly relevant because the last-invaded species and the newly invading species in a given region may belong to different taxonomic groups. With a similar goal, \citet{bonnamour2023historical} examine whether non-native plants facilitate alien insect invasions by providing suitable habitats and resources. They assess this by testing the predictive power of plant flows on non-native insect flows, aggregated over time. Despite using a different methodological approach, our findings align with theirs, showing that the strongest species invasion random effects involve a plant as the last-invaded species and an insect as the newly invading species. Furthermore, from a covariate perspective, the vast majority of last species–species interactions correspond to plant–insect relationships, such as between  \emph{Phenacoccus manihoti} and \emph{Chromolaena odorata}.

Our estimation method relies on the sampled partial likelihood, which at each event time considers a sampled risk set with only one event and one non-event. This reduces computational complexity, with only minimal information loss. In particular, the computational effort in a Cox Proportional Hazard Model dealing with splines and relying on the partial likelihood scales as $ O(n_\mathcal{S}  n_\mathcal{C}  n D^3) $, where $D$ is the dimension of the model matrix of the fitted GAM. Instead, the case-control partial likelihood via GAMs scales as  $ O(n D^3) $. The {\tt gamm} function in the {\tt mgcv} package is able to add random effects at the cost of 1 degree of freedom, but the disadvantage of this function is its computational cost. Instead, we use the 0-dimensional spline implementation in the {\tt gam} function in the same package, which is much faster, but which requires by default that $n>D$.  Due to this implementation constraint, we did not include all possible levels of the species co-invasion network.  Alternatively, it is possible to sample more than one non-event for each observed instance to increase $n$ and to avoid this artificial constraint. 


Modelling the high-quality, but binary \emph{first} records data meant that we only considered a binary counting process. Novel developments, including citizen science initiatives, are nowadays generating a richer, albeit more noisy and more complex, picture of species dispersion. Modelling the number of instances that a particular species was detected each year in each region could potentially improve the picture of the drivers of the dynamics of species invasions \citep{bonnamour2021insect}. 

\section{Competing interests}
None

\section{Author contributions statement}

\section{Acknowledgments}
This work was supported by funding from the Swiss National Science Foundation (grant 192549).

\section*{Data Availability and Reproducibility}

The data supporting this study are publicly available at \url{https://zenodo.org/record/4632335}. The code required to reproduce the analyses and results is accessible on GitHub at \url{https://github.com/martinaboschi/alienspecies.git}.

\bibliographystyle{plainnat}
\bibliography{reference.bib}

\section*{List of Figures}
\begin{itemize}
	\item[-] \textbf{Fig. 1.} \textbf{Simulation Study: Summary of the Results}. a) Comparison between the true coefficients (black solid) and the estimated coefficients for \textit{climatic dissimilarity} (yellow dashed) and \textit{distance} (blue dashed) on the simulated data. b) Comparison between the true (black solid) and the non-parametric estimates (blue dashed) of the cumulative baseline hazard. True baseline hazard is assumed to be constant and equal to $0.008$. c) Comparison between the true (black solid) and estimated random effects for species invasiveness (yellow) and region invasibility (blue). True values are represented by the conditional expectation of random effects fit on real data, while estimates correspond to the  $0$-dimensional spline estimates on the simulated data. Comparison has been performed in terms of empirical cumulative distribution function. d) Empirical distribution of the p-values resulting from testing \textit{climatic dissimilarity} (yellow), \textit{distance} (blue) and their \textit{global test} (brown). The empirical distributions are compared with the uniform cumulative distribution (black), which is the expected distribution of the p-values when the model is adequate.
	
	\item[-] \textbf{Fig. 2.} \textbf{Simulation Study on Recording Bias}. Estimated coefficients from 100 experiment replications are compared to the true parameters used in data simulation. The maximum masking probability represents the upper limit of region-specific masking probabilities, ranging from 0 to this maximum value. While increasing the maximum masking probability shrinks the estimates towards zero, the sign of the effect remains correctly identified.
	
	\item[-] \textbf{Fig. 3.} \textbf{Model Selection}. Values of AIC for the examined model formulations. We outline that, whenever included, \textit{distance}, \textit{trade}, and \textit{agricultural land-coverage} are supposed to have a time-varying impact while \textit{climatic dissimilarity}, \textit{urban land-coverage}, and \textit{colonial ties} a fixed effect. This choice comes from the previous studies on the topic \citep{Juozaitiene2022}. On the other side, all the considered models include random effects for species invasiveness, region invasibility and species co-invasion. The best model in terms of corrected AIC \citep{woodAIC} includes \textit{distance}, \textit{trade}, \textit{colonial ties} and \textit{climatic dissimilarity} and is outlined in the plot with a red crossed symbol. According to the covariates included in the compared models, we can distinguish seven groups of model formulations.

	\item[-] \textbf{Fig. 4.} \textbf{Time-varying estimates}. Time-varying estimated coefficients for a) \textit{distance} and b) \textit{trade} (undashed lines) with the related posterior confidence intervals (dashed lines).
	
	\item[-] \textbf{Fig. 5.} \textbf{Regions' invasibility in terms of estimated random effects}. Lightest-blue areas are those that the model identifies as most popular, such as Australia and Canada. On the other hand, darker locations are those that lead to a decrease in the rate of occurrence of alien species invasions (Peru and Saudi Arabia are some instances).

	\item[-] \textbf{Fig. 6.} \textbf{Strongest associations between species in terms of co-invasion}. We report here the values that are equal or larger, in absolute value, than the logarithm of $1.5$. Taking into account all the other features, the presence of the linked species-last species co-invasion lead to increasing or decreasing the risk of invasion at least of $ 50 \% $. The taxonomy determines the colour of the nodes (blue for plants and yellow for insects). When the estimated random coefficient is positive, the link is shown in black; a positive link indicates a higher probability of the sender node following the receiver node, meaning the risk for the sender species of entering a nation that has just been invaded by the recipient species in the plot is increased. The thicker the edge, the larger (in absolute value) the estimate for the corresponding random effect. The figures represent the nodes with the largest values in terms of species invasiveness.

	\item[-] \textbf{Fig. 7.} \textbf{Matched and Pooled Estimates of the Cumulative Baseline Hazard}. Following \citet{borgan1997risk}, we may provide a taxa-specific estimate of the baseline (blue for insects, yellow for plants, solid); as an alternative, we may rely on a pooled estimate (dotted) that takes into account the fact that the sampled control belongs to the same group as the observed case. b) \textbf{Goodness of Fit of Distance Covariate}. Graphical comparison between the observed squared norm of the martingale-residual process against $200$ simulated squared norm of Brownian Bridges. We plot the time in the original scale in order to understand which section of the time-window may lead to misspecification. It is interesting to see that the Bridge overcome the cloud created by the theoretical curves in the time range between $1961$ and $1983$ and in the time window between $1993$ and $2004$. We hypothesize that the impact of distance may have evolved in recent times, suggesting that more complex, non-linear, effects of distance, could improve the model fit.

\end{itemize}

\end{document}